\documentclass[usenatbib,twocolumn]{mn2e}
\usepackage{graphicx}
\usepackage{amsfonts}
\usepackage{epsfig,rotating,pstricks}
\usepackage{color}
\usepackage{amssymb,latexsym}
\usepackage{amsmath,wasysym}
\usepackage{verbatim}
\usepackage[T1]{fontenc}
\usepackage{aecompl}
\def\mhi{M_{\text{HI}}}
\def\w50{w_{50}}
\def\Vrot{V_{\text{rot}}}
\def\Vrothi{V_{\text{rot}}^{\text{HI}}}
\def\Vmax{V_{\text{max}}}
\def\Vmin{V_{\text{min}}}
\def\Vflat{V_{\text{flat}}}
\def\lsim{\apprle}
\def\gsim{\apprge}
\def\mstar{M_{\text{star}}}
\def\mhalo{M_{\text{h}}}
\def\msun{M_{\odot}}
\def\tage{t_{\text{age}}}
\def\beq{\begin{equation}}
\def\eeq{\end{equation}}
\def\phihi{\phi^{\text{HI}}}
\def\mhalostar{M_{\text{h}}^*}
\title[HI in DM Halos]{The Dark Matter Halos of HI Selected Galaxies}

\author[]{\parbox{18cm}{Saili Dutta$^{1}$\thanks{E-mail: sailidutta@niser.ac.in (SD)}, Nishikanta Khandai$^{1}$\thanks{E-mail: nkhandai@niser.ac.in (NK)}, 
Sandeep Rana$^{1}$\thanks{E-mail: sandeepranaiiser@gmail.com (SR)}}
  \vspace{0.3cm}\\
  $^{1}$ {School of Physical Sciences, National Institute of Science Education and Research, HBNI, Jatni 752050, India}\\
  }
\label{firstpage}

\def\LaTeX{L\kern-.36em\raise.3ex\hbox{a}\kern-.15em
    T\kern-.1667em\lower.7ex\hbox{E}\kern-.125emX}

\pagerange{\pageref{firstpage}--\pageref{lastpage}}

\begin{document}

\maketitle

\begin{abstract}

We present the neutral hydrogen mass ($\mhi$) function (HIMF) and
velocity width ($w_{50}$) function (HIWF)
based on a sample of 7857 galaxies from the
40\% data release of the ALFALFA survey ($\alpha.40$).
The low mass (velocity width) end of the HIMF (HIWF) is dominated by
the blue population of galaxies whereas the red population dominates
the HIMF (HIWF) at the high mass (velocity width) end.
We use a deconvolution method to estimate the
HI rotational velocity ($V_{\text{rot}}$) functions (HIVF) from the HIWF
for the total, red, and blue samples.
The HIWF and HIVF for the red and blue samples
are well separated at the knee of the function compared to their HIMFs.
We then use recent stacking results from the ALFALFA survey to constrain the halo mass ($\mhalo$)
function of HI-selected galaxies. This allows us to obtain various scaling relations
between $\mhi-\w50-\Vrot-\mhalo$, which we present.
The  $\mhi-\mhalo$ relation has a steep
slope $\sim 2.10$  at small masses and flattens to
$\sim 0.34$ at masses larger than a transition halo mass, 
$\log_{10}(M_{\text{ht}}h_{70}^2/M_{\odot})=10.62$.
Our scaling relation is robust and consistent with a volume-limited sample of $\alpha.40$.
The $\mhi-\mhalo$ relation is qualitatively similar to the $\mstar-\mhalo$ relation
but the transition halo mass is smaller by $\sim 1.4$ dex compared  
to that of the  $\mstar-\mhalo$ relation. Our results suggest that baryonic processes like
heating and feedback in larger mass halos suppress HI gas on a shorter time scale compared to
star-formation.

\end{abstract}



\section{Introduction}

Over the past few decades a number of surveys using ground based and space based telescopes, 
have mapped different parts of the Universe, both in terms of survey area and depth, 
covering the entire  
range of the electromagnetic spectrum. These surveys have shed light on how galaxies 
have formed and evolved across cosmic time in cosmologically 
significant volumes  \citep{2014ARA&A..52..415M,2014PhR...541...45C}. 
Of particular interest is to understand the relative abundance of various 
components of galaxies, e.g. stellar populations, gas -- both cold (e.g. molecular and atomic hydrogen)
and hot (as seen in X-ray observations of galaxy clusters) -- dust, metals and also  
supernovae and supermassive blackholes -- responsible for self-regulating the growth 
of galaxies through feedback processes. Targeted observations  are able to capture some 
of these individual properties of galaxies and allow us to look for correlations between them. These 
scaling relations give us some insight into galaxy formation and are also used as inputs 
for \emph{subgrid} prescriptions in theoretical models 
of galaxy formation, e.g. semi-analytical models \citep{2015ARA&A..53...51S}
and cosmological hydrodynamical simulations \citep{2017ARA&A..55...59N}.  
   
The common ingredient in both approaches is an $N-$body code which determines 
how initial density perturbations evolve to form large scale structures under gravitational 
instability, assuming a cosmological model which is now observationally constrained to better than
10\% \citep{2016A&A...594A..13P}. In the former, the entire merger tree of halos is used as a skeleton 
on which semi-analytical recipes for galaxy formation 
are assigned and model parameters are tuned to reproduce only \emph{certain} observations.
In this work (section~\ref{sec_scaling}) 
we will use one such galaxy catalog \citep{2018MNRAS.474.5206K} which 
is calibrated to reproduce the stellar mass function of galaxies at $z=0$. Cosmological 
hydrodynamical simulations, on the other hand, attempt to self-consistently evolve 
baryonic processes, so as to reproduce as many observational properties across cosmic time.
However since these simulations  are unable to resolve processes like star formation, 
supernovae explosions or gas accretion onto supermassive blackholes (to name a few) they  
resort to subgrid prescriptions which are calibrated from observations. Over the years, these 
simulations have refined their feedback models and are able to reproduce a number of observations, 
like the buildup of stellar mass and the star formation history of the Universe, 
stellar and quasar abundances and clustering  
and even scaling relations which have not been used as inputs in the simulations
\citep{2012ApJ...745L..29D,2014Natur.509..177V,2015MNRAS.446..521S,2015MNRAS.450.1349K,
  2015MNRAS.450.1937C,2016MNRAS.455.2778F,2018MNRAS.475..648P,2019MNRAS.486.2827D}.

It is only in recent years 
the attention has turned to reproducing and predicting (albeit qualitatively)
the distribution of cold gas (e.g. atomic HI and/or molecular $\text{H}_2$ hydrogen) 
in galaxies both in semi-analytical models \citep{2017MNRAS.465..111K,2020MNRAS.493.5434S} 
and hydrodynamical simulations
\citep{2017MNRAS.464.4204C,2018ApJ...866..135V,2019MNRAS.487.1529D,
  2020MNRAS.497..146D,2021MNRAS.502.3158S}. This has been achieved in simulations 
by post processing  simulation data after accounting 
for self-shielding \citep{2013MNRAS.431.2261R}. Observationally cold gas (HI+$\text{H}_2$)
is correlated to the star formation rate (SFR) -- a relation also known as the Kennicutt-Schmidt law
\citep{1959ApJ...129..243S,1963ApJ...137..758S,1989ApJ...344..685K,1998ApJ...498..541K} --  
and represents a reservoir of fuel for future star formation in galaxies. This relation also 
motivates the subgrid model \citep{2003MNRAS.339..289S} 
for star formation in  cosmological hydrodynamical simulations outlined earlier. 
The simulated predictions of cold gas are at a very nascent stage and have to be refined to 
predict, not only abundances, but also clustering and bivariate 
\citep{2001MNRAS.327.1249Z,2013ApJ...776...74L}
or conditional  mass functions 
of gas rich galaxies 
\citep[][hereafter D20, D21 respectively]{2020MNRAS.494.2664D,2021MNRAS.500L..37D}.  

Alternately there have been empirical and statistical approaches, 
like the Halo Abundance Matching (HAM)
technique \citep{2009ApJ...696..620C,2010ApJ...717..379B} applied to HI 
\citep{2011MNRAS.415.2580K,2017MNRAS.470..340P,2021MNRAS.506.3205S} and HI halo models 
\citep{2017ApJ...846...61G,2017MNRAS.469.2323P,2018MNRAS.479.1627P,2019MNRAS.486.5124O,
  2021MNRAS.503.4147P,2021arXiv210504570P} 
which have attempted to relate HI to dark matter halos based on the abundances and clustering
of HI selected galaxies.

In terms of detections, gas observations have steadily increased over the past decade 
\citep[for a compilation of results see][ and references therein]
{2013ARA&A..51..105C,2018MNRAS.473.1879R} 
but are still outnumbered by observations which target the stellar component of galaxies, e.g. optical,
UV or IR surveys. This missing link, especially at redshifts $z \gsim 0.1 $, 
between gas and star formation in galaxies needs to be bridged both 
by observations and theoretical models. However our understanding of the relation between cold gas 
and galaxy and halo environment is still limited and we need to observationally put as many 
constraints as possible. \cite{2012ApJ...756..113H,2020MNRAS.491.4843R,2020MNRAS.499.5656R} 
used data from HI, optical and UV surveys
and observationally constrained mutivariate HI-stellar scaling relations for HI selected galaxies. 
 
In the first part of this work we use a catalog of HI selected galaxies 
to obtain the bivariate (HI-mass - velocity width) abundance of these galaxies. We  use 
this result to obtain scaling relations between HI mass, 
velocity width and rotational velocity of these HI selected galaxies 
and also for different optically defined populations among them. We then use a recent stacking result 
of \cite{2020ApJ...894...92G} to define an HI selected halo mass function. This allows us to obtain
a relation between HI mass and halo mass for an HI selected sample and also obtain scaling relations
between halo mass and other HI properties like HI rotational velocities.

We assume a flat cosmology with ($\Omega_m, \Omega_{\Lambda}, h$) = ($0.3,0.7,0.7$), 
where $h$ is the dimensionless Hubble parameter related 
to the Hubble constant $H_0 = 100 \, h \, \text{km.s}^{-1}\text{Mpc}^{-1}$. We follow the notation 
of \cite{2010MNRAS.403.1969Z} to define a normalized 
Hubble constant, $h_{70} = h/0.7 = H_0/(70 \, \text{km.s}^{-1}\text{Mpc}^{-1})$ 
which takes the value of 1 for our choice of $h=0.7$.

Our paper is organized as follows. In section~\ref{sec_data} we describe our data and sample. 
In section~\ref{sec_2dswml} we describe a likelihood 
method used to estimate the HI velocity width function
(HIWF). We outline a deconvolution method to obtain the HI velocity function (HIVF) 
from the HIWF in section~\ref{sec_hiwf} 
and present our results for the HIWF and HIVF for the full, red and 
blue samples of HI selected galaxies. In section~\ref{sec_scaling} we present scaling relations
between HI properties and between HI and halo properties. We discuss and summarize 
our results in section~\ref{sec_summary}.

\section{Data}
\label{sec_data}

The Arecibo Legacy Fast ALFA (ALFALFA) is a blind extragalactic HI survey,
detecting galaxies with HI masses ranging from $10^6$ to $10^{11}$ solar masses 
out to redshift $z=0.06$. 
The 40\% data release \citep{2011AJ....142..170H} 
of ALFALFA ($\alpha.40$) is a catalog of  
15855 galaxies, detected in the region
$7^h 30^m<$R.A.$<16^h 30^m$, 
$4^{\circ}<$ dec.$<16^{\circ}$, and 
$24^{\circ}<$ dec.$<28^{\circ}$ and
$22^h<$R.A.$<3^h$, $14^{\circ}<$ dec.$<16^{\circ}$, and
$24^{\circ}<$ dec.$<32^{\circ}$.
The $\alpha.40$ survey area is $\sim 2752$ deg$^2$, 
which is 40\% of the total targeted area.
Most of the galaxies from ALFALFA have been crossmatched with 
Sloan Digital Sky Survey (SDSS) data release 7 (DR7) 
\citep{2009ApJS..182..543A}.
Along with the observational properties like angular 
position (right ascension (RA), declination (dec)), 
heliocentric velocity ($cz_{\text{helio}}$), 
observed velocity profile width ($w_{50}$), integrated flux ($S_{21}$);
some derived properties, e.g.  
HI mass ($\mhi$), distance ($D$)
are also tabulated in the catalog. 
The catalog also includes a flag which classifies the detection quality,
based on the signal to noise ratio (S/N).
Code 1 objects ($N_{\text{gal}} = 11941$) are the best detections with $S/N > 6.5$.
Detections with $S/N < 6.5$ are referred to as code 2 objects ($N_{\text{gal}} = 3100$).
The catalog contains some High Velocity Clouds (HVC), 
denoted by code 9 objects ($N_{\text{HVC}} = 814$). \\

We consider only the code 1 objects for our analysis
and restrict our sample to $cz_{\text{cmb}} = 15000\; \text{km.s}^{-1}$ 
to avoid Radio Frequency Interference (RFI)
\citep{2010ApJ...723.1359M, 2011AJ....142..170H}, where $z_{\text{cmb}}$
is the galaxy redshift in the CMB reference frame. 
In this work we consider an area ($\sim 2093$ deg$^2$)
common to both ALFALFA and SDSS (as discussed in D20),
which contains $8344$ HI-selected galaxies.\\

From the observed distribution of HI-selected galaxies in the $S_{21}-w_{50}$ plane
\citep[][D20]{2011AJ....142..170H} we  see that at fixed integrated flux, $S_{21}$,  
galaxies having narrower velocity width profiles are more likely 
to be detected. The sensitivity limit depends therefore on both 
flux and velocity width. 
We apply the sensitivity limit given by the 50\% completeness 
limit \citep[see eqs.4-5 of][]{2011AJ....142..170H}, 
which reduces the sample to 7857 galaxies. 
Among these, 148 galaxies (referred to as \textit{dark galaxies}) 
do not have any optical counterparts 
in SDSS DR7 although they belong to the DR7 footprint.
We exclude these galaxies in this work as they will not affect our results (see D20). 
For the remaining 7709 galaxies we extract the \textit{ugriz} magnitudes from SDSS
(extinction corrected -- our galaxy) 
and \textit{kcorrect} \citep{2007AJ....133..734B} them to obtain 
rest frame magnitudes. \textit{kcorrect} also estimates stellar masses  
which we will use later in this work. We obtain galaxy age estimates based on  
the Granada Flexible Stellar Population Synthesis (FSPS) models 
\citep{2009ApJ...699..486C,2014ApJS..211...17A} from SDSS. We point out that while using the estimates from 
\textit{kcorrect} we have not corrected for internal reddening due to dust, 
but as shown in D20 our results do not change considerably with those of 
\cite{2012ApJ...756..113H}, who have corrected for reddening by using two 
additional UV bands of GALEX (Galaxy Evolution Explorer). 

A clear bimodality is seen in the distribution of SDSS galaxies 
in the color ($u-r$) - magnitude ($M_r$) plane. 
We use an optimal divider \citep{2004ApJ...600..681B} to 
further classify our sample as red or blue galaxies.
Red(blue) galaxies lie above(below) 
the optimal divider defined in the color-magnitude plane as
\begin{eqnarray} 
C^\prime_{ur}(M_r) &=& 2.06 - 0.244 \tanh \left[ \frac{M_r+20.07}{1.09}\right]
\label{eq_tanh_cut}
\end{eqnarray}
We have $N_{\text{gal}}^{red} = 1290 $ red galaxies 
and $N_{\text{gal}}^{blue} = 6419 $ blue galaxies in our observed sample. 

In this work we restrict ourselves with the $\alpha.40$ sample rather than the recently 
released 100\% catalog ($\alpha.100$) \citep{2018ApJ...861...49H,2020AJ....160..271D}. 
We find that a number of galaxies in ALFALFA with optical counterparts as seen in the SDSS images,
have been masked out due to bright foreground stars. Additional work will be required to extract 
the photometric properties of such galaxies which we will address in the future.

\section{Estimating the HI Velocity Width Function}
\label{sec_2dswml}
Similar to the HI mass function (HIMF), the  
HI velocity width function (HIWF) can be defined as
the underlying number density of galaxies with velocity widths 
in the range [$w_{50},w_{50}+dw_{50}$],
\begin{equation}
\phi(w_{50}) = \frac{1}{V} \frac{dN}{dw_{50}}
\end{equation}
where, $dN$ is the total number of galaxies in volume $V$
having velocity widths within $w_{50}$ and $(w_{50}+dw_{50})$.
The HIWF is well described by a modified Schechter function 
\citep{2010MNRAS.403.1969Z,2011ApJ...739...38P,2014MNRAS.444.3559M}
\begin{eqnarray}
\phi(w_{50}) &=& \frac{dn}{d\log_{10} w_{50}} \nonumber \\
&=& \ln (10) \phi_* \left( \frac{w_{50}}{w_*}\right)^{\alpha} 
\exp\left[-\left( \frac{w_{50}}{w_*}\right)^{\beta}\right]
\label{eq_modified_Schechter}
\end{eqnarray}
where $\phi_*$ is the amplitude, $\alpha$ is the slope at the low velocity width end, 
$w_*$ is the characteristic velocity width, or the knee of the Schechter function, 
and $\beta$ modifies the exponential suppression at high velocity widths. 
For the rest of the paper we will quote the amplitude of the Schechter function, $\phi_*$, 
and the knee of the HIWF, $w_*$, in units of 
$(h_{70}^{3} \text{Mpc}^{-3} \text{dex}^{-1})$ and 
$\log_{10}[w_*/(\text{km.s}^{-1})]$ respectively. Later on we will also look at the HI 
velocity function (HIVF), described by a modified Schechter function whose 
knee, $V_*$, will similarly be in the same units as $w_*$. Finally the values of the 
HI mass, $\mhi$, (and the knee of the HIMF, $M_*$)  will be in units of
$\log (\mhi/M_{\odot}) + 2\log h_{70}$. 
Similarly, the values of halo mass, $M_{\text{h}}$, and 
stellar mass, $M_{\text{star}}$, will be in units of 
$\log (M_{\text{h}}/M_{\odot}) + 2\log h_{70}$ and 
$\log (M_{\text{star}}/M_{\odot}) + 2\log h_{70}$, respectively, throughout the paper.

\begin{figure*}
\centering
  \begin{tabular}{cc}
    \includegraphics[width=3.4in]{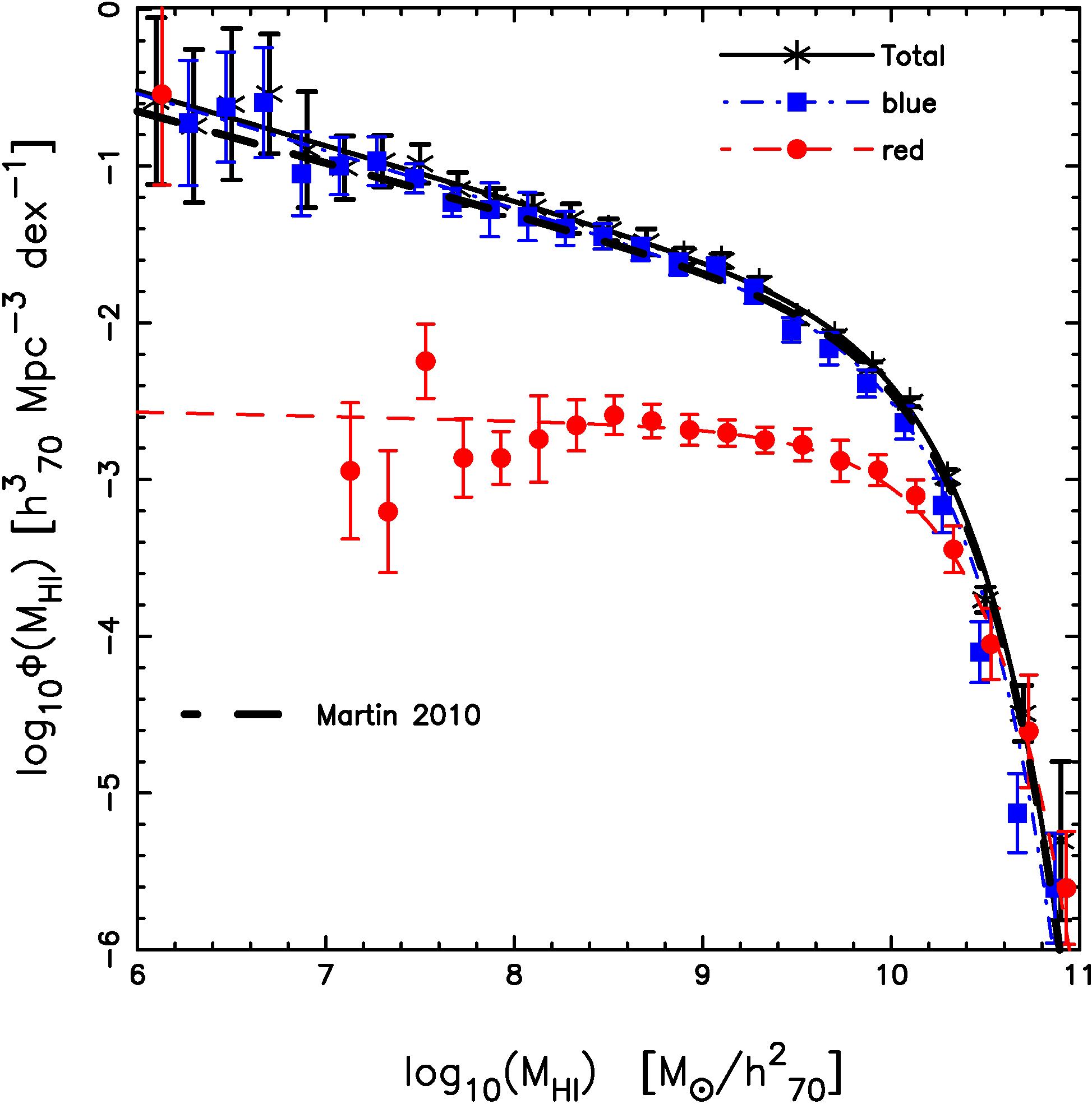} 
    \includegraphics[width=3.4in]{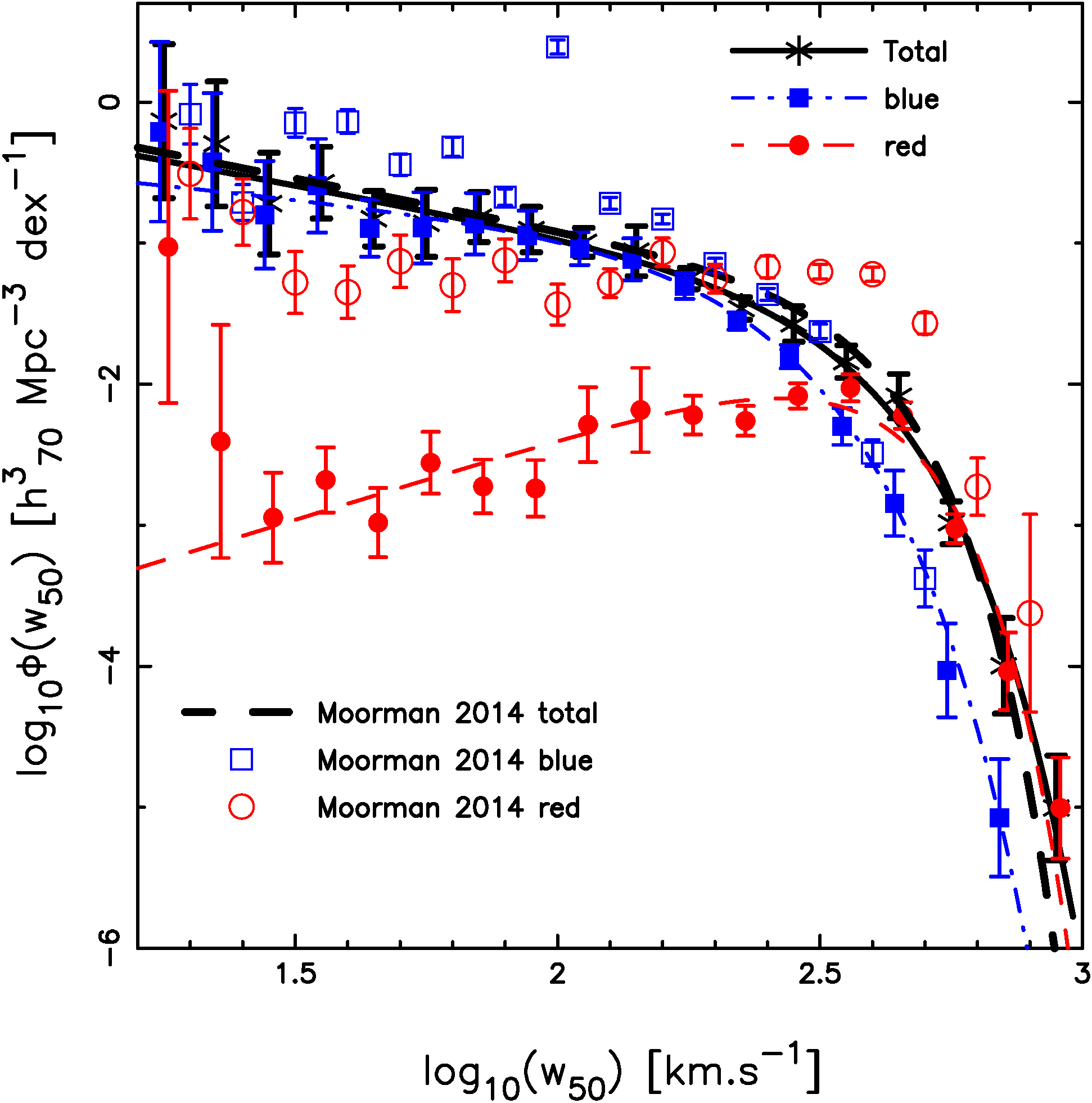}\\
  \end{tabular}
  \caption{Our estimates of the HIMFs (HIWFs) are 
shown in the left (right) panel for the 
total (cross), blue (filled square) and red (filled circle) samples.
To better illustrate our results the data for the red (blue) sample 
have been horizontally offset rightward (leftward) with respect to the total sample.
The Schechter (modified Schechter) function fits are plotted 
for the total (thick solid line), red (thin dashed line) and blue (thin dot-dashed line)
samples for the HIMFs (HIWFs). 
The thick dashed line in the left panel 
is the estimate of the HIMF from \citet{2010ApJ...723.1359M}.
The thick dashed line in the right panel 
is the estimate of the HIWF from \citet{2014MNRAS.444.3559M}.
Both of these have used the $\alpha$.40 sample.
The open circles (squares) in the right panel are the unnormalized estimates of the HIWF
for the red (blue) sample from \citet{2014MNRAS.444.3559M}}
  \label{mf_wf_combo}
\end{figure*}

The step-wise maximum likelihood (SWML) \citep{1988MNRAS.232..431E} is used 
to estimate the underlying velocity width function. This model-independent method 
does not assume any functional form of the width function, rather it
estimates a discretized (or binned) width function, $\phi(w_{50}^k)$.
Being a maximum likelihood method, it is insensitive to the effects 
of local variations in galaxy densities due to clustering.
For HI-selected galaxies the detection probability depends both on $\mhi$ and $\w50$.
We implement a two-dimensional SWML method (2DSWML) to first estimate the bivariate
HI mass-velocity width function, $\phi(\mhi,\w50)$ and then integrate over $\mhi$ to obtain 
the HIWF. The approach is similar to 
\citet{2000MNRAS.312..557L,2003AJ....125.2842Z, 2010ApJ...723.1359M, 2011AJ....142..170H} 
and has been described in detail in D20 in the context of obtaining the HIMF. 
We briefly outline the method below.

To estimate the bivariate distribution function $\phi_{jk} \equiv \phi(\mhi^j,\w50^k)$,
we divide the HI mass range and HI velocity width range into logarithmic 
$n_M$ and $n_W$ bins,
where $j=0,1,..,n_M-1$ and $k=0,1,..,n_W-1$.
The probability of detecting a galaxy \emph{i} of HI mass $M^i_{\text{HI}}$ and 
HI velocity width $w^i_{50}$ at a distance $D^i$ is
\begin{eqnarray}
p_i &=& \frac{\Sigma_j \Sigma_k V_{ijk} \phi_{jk}}
{\Sigma_j \Sigma_k H_{ijk} \phi_{jk} \Delta M \Delta W}
\label{eq_probi}
\end{eqnarray}
where $\Delta M$ and $\Delta W$ are the bin widths in the 
mass $M=\log_{10}[\mhi/\msun]$ and velocity width $W=\log_{10}[\w50/(\text{km.s}^-1)]$ 
axes respectively.
$V_{ijk}$ is an occupation number and takes a value of 1 (0 otherwise) 
if  the galaxy \emph{i} is binned in the '$jk$' bin, 
$\Sigma_i V_{ijk} = n_{jk}$,
where $n_{jk}$ is the observed number of galaxies in '$jk$' bin.
$H_{ijk}$ takes care of the completeness limit of the sample and
represents the fractional area in the $\mhi-\w50$ plane that is accessible to the 
\emph{i}$^{th}$ galaxy. It takes values from 0 to 1 \citep[See][]{2020MNRAS.494.2664D}.
The joint likelihood, $\mathcal{L} = \prod_{i=1}^{N_{\text{gal}}} p_i$ is then 
maximized with respect to $\phi_{jk}$,
which gives us an expression for $\phi_{jk}$
\begin{eqnarray} \label{eq_phijk}
\phi_{jk} &=& n_{jk}\left[\Sigma_i \frac{H_{ijk}}
{\Sigma_m \Sigma_n H_{imn} \phi_{mn}}\right]^{-1}
\end{eqnarray}
which needs to be solved iteratively. 
The iteration is started by setting $\phi_{jk}$ to $n_{jk}/V$ and is stopped when we achieve
a minimum 1\%  accuracy for all $\phi_{jk}$.   
Finally we integrate the bivariate $\phi_{jk}$ 
over $M_{\text{HI}}$ to obtain the HIWF:
\begin{eqnarray}
\phi_k &=& \sum_j \phi_{jk} \Delta M
\end{eqnarray}
Integrating $\phi_{jk}$ over $\w50$, on the other hand, gives us the HIMF $\phi_j$.
A common feature of maximum likelihood methods is that the normalization 
of $\phi(\mhi,\w50)$ needs to be fixed separately since  it gets lost in the process (see 
equation~\ref{eq_probi}). 
We do this by computing the selection function and matching the observed number density 
to the underlying number density convolved by the selection function 
\citep{1982ApJ...254..437D, 2010ApJ...723.1359M}.

\subsection{Uncertainties on HIWF}
We consider the following four sources of error 
and add them in quadrature to quantify 
the uncertainty on the estimated HIWF, similar to \cite{2014MNRAS.444.3559M}.

\begin{enumerate}
\item \textbf{Velocity width errors:} 
Errors due to the measurement of $w_{50}$ can change 
the occupation of one galaxy in $M_{\text{HI}}-w_{50}$ plane.
To account for this we have considered 300 realizations of $w_{50}$
(Gaussian random realizations using $w_{50}$ as the mean 
and $\sigma_{\w50}$ as the variance)
and estimated HIWF for each of these realizations. 
The distribution of $\phi_k$ for a fixed \emph{k} 
gives an estimate of the uncertainty on $\phi_k$.

\item \textbf{Distance errors:} 
HI mass ($\mhi$) depends on distance, and since we estimate the HIWF 
from a bivariate function of $M_{\text{HI}}$ \& $w_{50}$,
the measurement errors on distance is also a source of uncertainty.
We follow a similar approach as above to get the errors on $\phi_k$, 
by generating 300 random realizations.

\item \textbf{Sample variance:} 
We split our survey area into 26 regions of approximately equal angular area
and estimate the HIWF by eliminating one region at one time (the Jackknife sample).
We then estimate the Jackknife uncertainty as 
$\sigma_{\phi_k} = \frac{N-1}{N}\sum_{i=1}^{N} 
(\bar{{\phi}_k} - \phi^i_k)^2$ where $N$ is the number of Jackknife samples,
$\bar{{\phi}_k}$ is the Jackknife mean and 
$\phi^i_k$ is the value for the $i^{\text{th}}$ Jackknife sample.

\item \textbf{Poisson errors:} 
The observed counts at the two ends of $\phi(\w50)$ are small, it is therefore 
important to consider Poisson errors.
\end{enumerate}

\section{HI Velocity Width Function}
\label{sec_hiwf}

\begin{figure*}
\centering
  \begin{tabular}{cc}
    \includegraphics[width=3.4in]{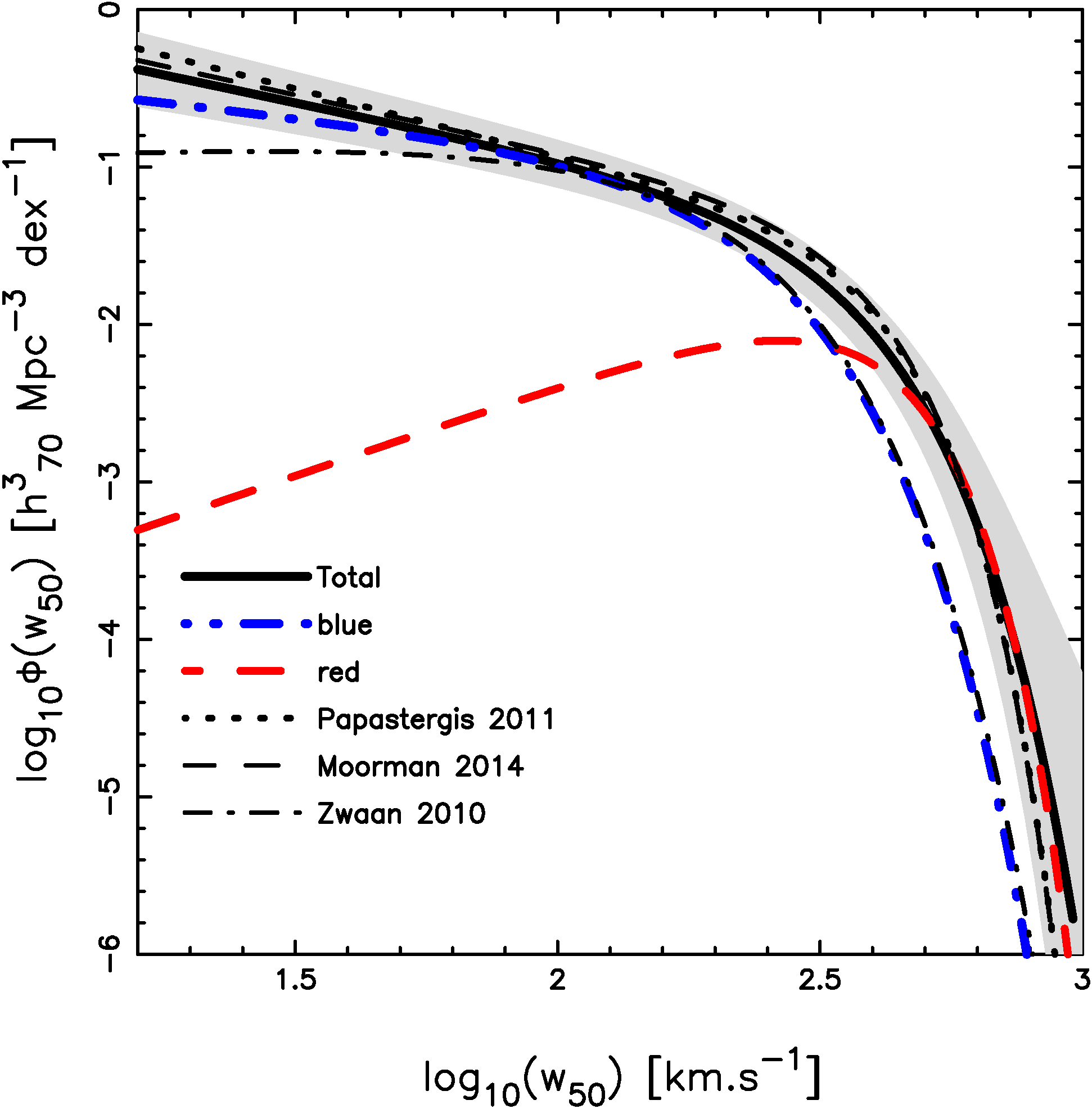} 
    \includegraphics[width=3.4in]{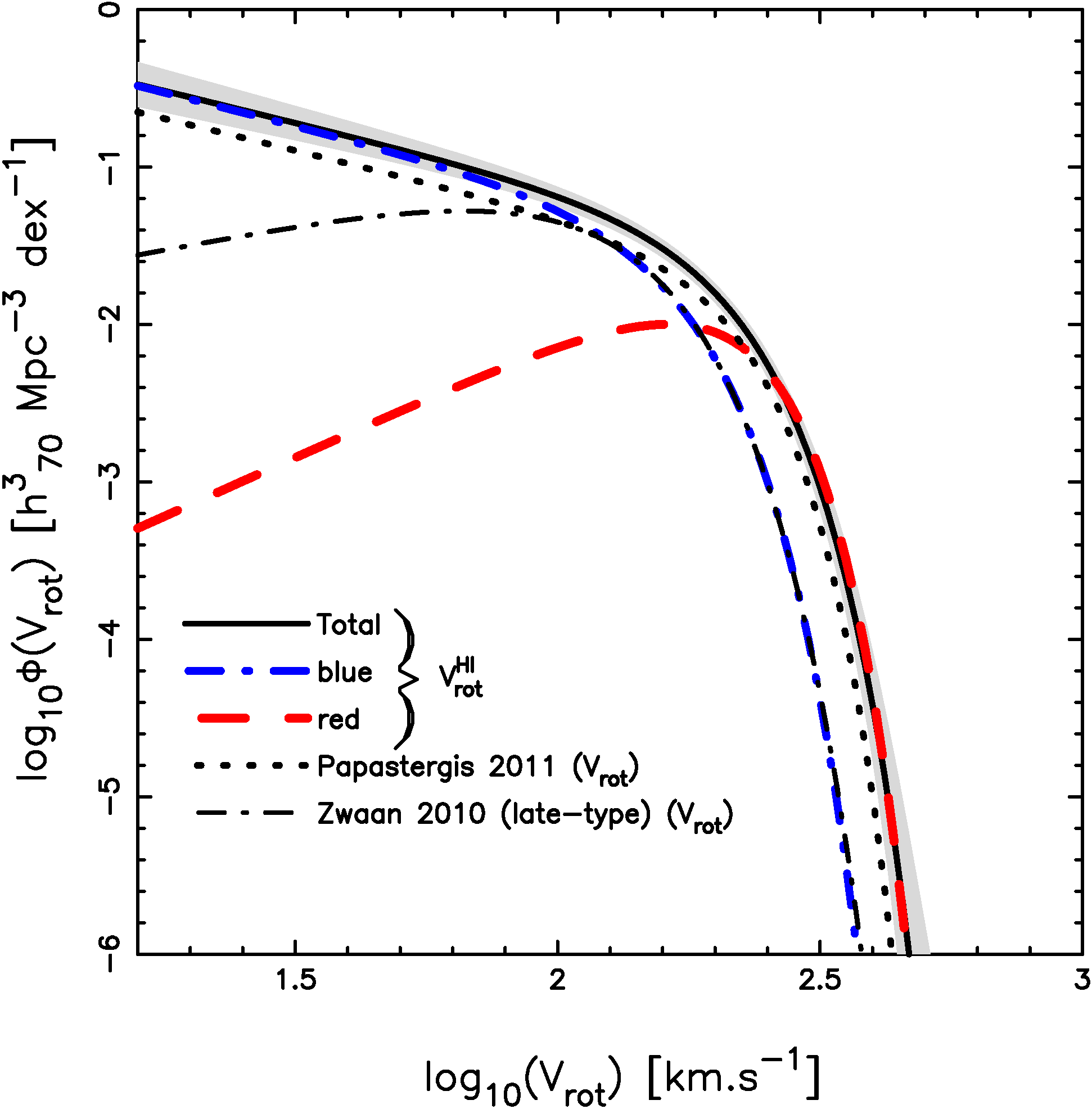}\\
  \end{tabular}
  \caption{A comparison of our estimates with previously published results
    of the HIWF (HIVF) are shown in the left (right) panel. 
    Our modified Schechter function fits and their $1\sigma$ uncertainty 
    (shaded region for the total sample only)
    are plotted for the total (thick solid line), red (thin dashed line) 
    and blue (thin dot-dashed line) samples in both panels. 
    The thick dotted line represents the estimates of \citet{2011ApJ...739...38P}
    for both the HIWF and the HIVF. The thin dot-dashed lines are the estimates of 
    \citet{2010MNRAS.403.1969Z} from the HIPASS data for the full sample (left panel, HIWF)
    and for late-type galaxies (right panel, HIVF).
    The thin dashed line is the estimate of the HIWF of \citet{2014MNRAS.444.3559M}.
  }
  \label{wf_vf_combo}
\end{figure*}

In this section we present the HIMF and the HIWF for the full sample as well as for the 
red and blue samples. We then use a deconvolution (or inversion) method \citep{2011ApJ...739...38P}
to estimate the HIVF from the HIWF. 

In figure~\ref{mf_wf_combo} we present our estimates of the HIMF (left panel) 
and the HIWF (right panel) for the total (cross), red (filled circle) and 
blue (filled square) samples. Error estimates on the HIMF are described in D20 
and are similar to the error estimates for the HIWF. To better display our data  
we have horizontally offset data for the red and blue samples with respect 
to the total sample. The thick solid, thin dashed and thin dot-dashed lines are 
our Schechter (modified Schechter) function fits to the HIMF (HIWF) for the total, 
red and blue samples. We compare our results 
for the HIMF (HIWF) with \cite{2010ApJ...723.1359M} \citep{2014MNRAS.444.3559M} 
(thick dashed line). We find that our estimates compare well with 
these authors for the total sample.
\cite{2014MNRAS.444.3559M} also estimated the HIWF for the red
(open circles) and blue (open square) samples but they are most likely unnormalized 
since they do not add up to give the HIWF for the total sample. The overall shape 
of the red sample compares well with \cite{2014MNRAS.444.3559M}; however, since the 
binning is different, it is difficult to make a point by point comparison. 
\cite{2014MNRAS.444.3559M}  have an outlier at $\w50 \approx 100 \text{km.s}^{-1}$. 
We find that it comes from a single galaxy which is extremely gas poor 
with $\mhi = 6.78$. We have removed this object in our analysis.  

Galaxies have random inclinations, therefore 
we relate the observed HI profile width, $\w50$, to the intrinsic HI rotational 
velocity of the galaxy, $\Vrothi$, by 
\beq
\w50 = 2 \Vrothi \sin{i} + w_{\text{nr}}
\label{eq_w50_vrot}
\eeq 
This is similar to the approach taken by 
\cite{2010MNRAS.403.1969Z,2011ApJ...739...38P}. Here $i$ is the inclination angle
and $w_{\text{nr}}$ is an additional term which captures broadening by turbulence 
and other non-rotational motion. We will discuss later the use 
of eq.~\ref{eq_w50_vrot} which is valid when $w_{\text{nr}} \ll \Vrothi$.

\begin{table*}
\begin{center}
\begin{tabular}{|l|c|c|c|c|c|c|c|c|c|}
\hline
&  & $\phi(w_{50})$ & & & & & $\phi(\Vrothi)$ & &   \\
\hline
&  $\phi_*$ & $w_*$ & $\alpha$ & $\beta$ & & $\phi_*$ & $V_*$ & $\alpha$ & $\beta$ \\
\hline
total & 0.023 $\pm$ 0.008 & 2.50 $\pm$ 0.06 & -0.70 $\pm$ 0.16 & 2.02 $\pm$ 0.23  & & 0.0187 $\pm$ 0.0023 & 2.3 $\pm$ 0.01 & -0.81 $\pm$ 0.13  & 2.7 $\pm$ 0.17\\
\hline
blue & 0.046 $\pm$ 0.014 & 2.29 $\pm$ 0.06 & -0.37 $\pm$ 0.18 & 1.74 $\pm$ 0.16 & & 0.0206 $\pm$ 0.0023 & 2.2 $\pm$ 0.01  & -0.84 $\pm$ 0.16  & 2.7 $\pm$ 0.18\\ 
\hline
red  & 0.008 $\pm$ 0.002 & 2.56 $\pm$ 0.07 & 1.14 $\pm$ 0.29 & 2.51 $\pm$ 0.45 & & 0.0102 $\pm$ 0.0013 & 2.31 $\pm$ 0.01 & 1.5 $\pm$ 0.18 & 2.96 $\pm$ 0.14  \\ 
\hline
\end{tabular}
\end{center}
\caption{Best fit values of modified Schechter parameters for 
$\phi(\w50)$ and $\phi(\Vrothi)$ for total, blue and red samples.}
\label{tab_modified_schechter}
\end{table*}

Although it is tempting to relate $\Vrothi$ to rotational velocities, $\Vrot$, 
associated with rotation curve measurements by radio synthesis observations, one needs to be careful
in this regard. \cite{2001ApJ...563..694V} classified rotation curves 
broadly as i. rising (R-type) 
ii. flat (F-type) and iii. declining (D-type) rotation curves.  R-type rotation curves 
\citep{2020ApJS..247...31L} are 
generally associated with low-surface brightness or dwarf galaxies and the 
observed maximum rotational velocity, $\Vmax$,  
is  the last point of the rotation curve measurement. It represents a lower bound 
to the rotational velocity, $\Vrot$, associated with the gravitational potential.  
F-type rotation curves \citep{2020ApJS..247...31L} 
are the classical flat rotation curves that extend beyond 
the optical radius. The rotation curves initially rise, reach a maximum value $\Vmax$, and 
taper off to a slightly lower, constant value, $\Vflat \lsim \Vmax$,  
at large radii. $\Vflat$ is associated 
with the maximum rotational velocity, $\Vrot$, that would be induced by the mass of the 
halo. F-type curves are associated 
with late-type spirals and $\Vrothi$, in equation~\ref{eq_w50_vrot}, 
can be associated with $\Vflat = \Vrot = \Vrothi$ \citep{2001ApJ...563..694V}.  
Finally the D-type curves \citep{2017ApJ...836..152L}, 
seen in early-type galaxies (ETGs), 
are characterized by an increasing rotation curve reaching to $\Vmax$ within the stellar disk
and declining thereafter beyond the optical radius to either a constant value,$\Vflat$,
or to a smaller value, $\Vmin$. In the D-type curves, $\Vmax > \Vflat,\Vmin$. 

Among the small fraction of ETGs detected in HI, 
in the ATLAS$^{\text{3D}}$ survey \citep{2012MNRAS.422.1835S} 
most have settled HI disk configurations which are rotating. However while creating dynamical 
mass models based on the rotation curves \citep{2016AJ....152..157L,2017ApJ...836..152L}
one has to also consider stellar velocity dispersion and pressure support of ionized X-ray gas 
in such systems \citep{2017ApJ...836..152L}. The corresponding $\Vrot$ based on these models
is not the same as $\Vrothi$. Pressure support is also important in dwarfs described 
by R-type rotation curves.

Given these considerations we will interpret $\Vrothi$ in equation~\ref{eq_w50_vrot} 
as the HI profile width if the HI disk were seen edge-on, corrected for non-rotational broadening.  
In figure~\ref{fig_mhi-vrot} we see that the $\mhi$-$\Vrothi$ 
relation in our sample is broadly consistent with the $\mhi$-$\Vflat$ measurements 
from the Spitzer Photometry and Accurate Rotation Curves (SPARC)\citep{2016AJ....152..157L} sample.
   
Broadly there are two approaches in obtaining $\Vrothi$ from $\w50$. One can correct 
for inclination effects \citep{2010MNRAS.403.1969Z} by estimating inclination angles from identified 
optical counterparts. This method however is accurate for inclination angles $i \leq 45^o$. 
Because of the ambiguity of interpreting  $\Vrothi$ with $\Vrot$ for all morphological types,
\cite{2010MNRAS.403.1969Z} estimated the HI velocity function (HIVF) with the 2DSWML method 
only for late-type galaxies with this method, 
after  applying a correction in the abundance for galaxies 
with inclination angles $i > 45^o$. 
Alternately the rotation curve $V_{\text{rot}}(r)$, and the surface 
density of the HI disk, $\Sigma_{\text{HI}}(r)$,  
can produce the observed HI flux density profile, $S_{\text{HI}}(v)$, 
after convolving with the telescope beam response and non-rotational broadening (arising due to 
turbulent motions in the disk) and accounting for inclination 
\citep[][]{1971ApJ...169..235G,1994ApJ...423..180S,2021arXiv210504570P}. As a proof of concept,
\cite{2021arXiv210504570P} turned this expression (see their eq.3) 
around to infer halo properties from the observed 21cm line profile, $S_{\text{HI}}(v)$, 
of two galaxies NGC-99 and UGC-00094 in ALFALFA, 
by considering a baryonification model of halos to produce rotation 
curves \citep{2021MNRAS.503.4147P} and using an observationally constrained 
scaling of the HI scale length to HI mass,  
$h_{\text{HI}} \propto \mhi^{0.5}$ \citep{2016MNRAS.460.2143W}, 
which determines    $\Sigma_{\text{HI}}(r) \propto \exp(-r/h_{\text{HI}})$. 
\cite{2021arXiv210504570P} also showed that the mass weighted second moment of the rotational
velocity is related to $w_{50}$ using this relation. Although there are many assumptions and parameters 
which lead us from halos to the HI line profile, this method provides a novel way to estimate
the rotational velocity from the observed 21cm spectrum $S_{\text{HI}}(v)$. One can  then use 
the 2DSWML method to estimate the rotational velocity function (mass weighted) of HI selected galaxies
as was done by \cite{2010MNRAS.403.1969Z}.   

The second approach is a statistical method  implemented 
by \cite{2011ApJ...739...38P}. We follow this approach. We assume that the HIVF is also 
a modified Schechter function described by equation~\ref{eq_modified_Schechter} with  parameters 
($\phi_*, V_*, \alpha, \beta$). We generate a realization of this 
model Schechter function and randomize 
their inclinations ($\cos{i}$ uniformly distributed) and then use equation~\ref{eq_w50_vrot} 
to obtain a realization of the HIWF. We use $w_{\text{nr}} = 5 \text{km.s}^{-1}$ 
in equation~\ref{eq_w50_vrot} 
\citep{2001A&A...370..765V,2011ApJ...739...38P}, and add it linearly for galaxies 
with $\Vrot > 50 \text{km.s}^{-1}$ and in quadrature for galaxies with smaller velocities 
\citep{2011ApJ...739...38P}.
 
Although \citet{2013ApJ...773...88S} refer to a bit higher values ($5 - 15 \; \text{km.s}^{-1}$) for $w_{nr}$, we find negligible differences in the derived HIVFs for different $w_{nr}$ 
values and stick with $w_{\text{nr}} = 5 \text{km.s}^{-1}$ 
in order to be consistent with \cite{2011ApJ...739...38P}. We point
the reader to appendix~\ref{appendix} for a more detailed discussion on this issue.

The model HIWF is compared to our estimated binned HIWF
(data points with errors in figure~\ref{mf_wf_combo}) and a $\chi^2$ is computed using 
the model, data and associated errors. Finally the best fit model parameters of the HIVF 
are obtained by minimizing the $\chi^2$ with the model parameters. We do this for the total, red
and blue samples.

Our results are shown in figure~\ref{wf_vf_combo} and summarized in table~\ref{tab_modified_schechter}.
In figure~\ref{wf_vf_combo} we compare our modified Schechter function fits with those obtained 
earlier with the HI Parkes All Sky Survey, \citep[HIPASS,][]{2004MNRAS.350.1195M,2010MNRAS.403.1969Z} 
and the $\alpha$.40 sample of ALFALFA 
\citep{2011ApJ...739...38P,2014MNRAS.444.3559M}. The thick solid, thick dashed and thick dot-dashed 
lines are our modified Schechter function fits for the total, red and blue samples. The grey region 
is the $1\sigma$ uncertainty for the total sample. The dotted line is the result of 
\cite{2011ApJ...739...38P}. The thin dashed line in the left panel of figure~\ref{wf_vf_combo}
is the estimate of the HIWF by \cite{2014MNRAS.444.3559M}. The thin dot-dashed line is the result of 
\cite{2010MNRAS.403.1969Z} for the HIWF (left panel, all galaxies) and the HIVF(right panel, 
late-type galaxies) in HIPASS.

As can be seen in figure~\ref{wf_vf_combo}, our result for the HIWF agrees well (within errors) with 
\cite{2011ApJ...739...38P,2014MNRAS.444.3559M}. There is a factor of $\sim 2$ 
discrepancy in $\phi_*$ between \cite{2011ApJ...739...38P} ($\phi_* = 0.011 \pm 0.002$) 
and \cite{2014MNRAS.444.3559M} ($\phi_* = 0.021 \pm 0.002$). Our value of 
$\phi_* = 0.023 \pm 0.008$ is in better agreement with \cite{2014MNRAS.444.3559M}. However given the 
correlation between the parameters of the Schechter function the  differences in other parameters 
compensate for each other and keep the overall shape within the $1\sigma$ uncertainty.  
The difference between the ALFALFA \citep[][this work]{2011ApJ...739...38P,2014MNRAS.444.3559M} 
results and HIPASS \citep{2010MNRAS.403.1969Z} are much starker. Although the full HIPASS sample 
has been taken to estimate the HIWF, the results  from
\cite{2010MNRAS.403.1969Z} seems to match only with the HIWF of blue galaxies in ALFALFA 
at the high velocity end, and has a shallower slope at lower velocity widths. The reason is that due to 
its higher sensitivity ALFALFA is able to detect larger velocity widths compared to HIPASS. 
As is evident from the figure, the large velocity end is dominated by red galaxies which 
one can associate with early-type galaxies (see figure~\ref{fig_age}), HIPASS would therefore miss
this population altogether. The observed fraction of early type galaxies in HIPASS is $11\%$
compared to $20\%$ for red galaxies in our sample. 

In the right panel of figure~\ref{wf_vf_combo} we show our estimates of the HIVF for the total, 
red and blue samples. Comparison is made with the total sample of ALFALFA 
\citep{2011ApJ...739...38P} and the late-type
galaxies in HIPASS \citep{2010MNRAS.403.1969Z}. 
Similar to what we saw in the HIWF, the high velocity end 
of the HIVF for the late type galaxies in HIPASS compares well with that of 
blue galaxies in our sample.
This suggests that associating $\Vrothi$ with $\Vrot$ is correct for late type galaxies. 
This also suggests that associating blue galaxies with late type galaxies with the age criterion given 
in figure~\ref{fig_age} is reasonable.   
Finally it confirms our earlier argument that HIPASS galaxies sample the blue cloud more 
frequently as compared to ALFALFA. 

For the total sample our results of the HIVF is systematically offset
compared to \cite{2011ApJ...739...38P} although our HIWF agree with each other. 
However \cite{2011ApJ...739...38P} have estimated 
the rotational velocity , $\Vrot$, function for HI-selected galaxies, whereas we have estimated
the HI rotational velocity, $\Vrothi$, function. To estimate the rotational velocity function 
\cite{2011ApJ...739...38P} estimated the HIVF for late type galaxies, based on the inversion method 
outlined above. For early type galaxies they used the velocity dispersion function, 
$\phi^{\text{early}}(\sigma)$, 
of early type galaxies obtained from SDSS and the 2dFGRS \citep{2010MNRAS.402.2031C}. 
$\phi^{\text{early}}(\sigma)$ was then converted to $\phi^{\text{early}}(\Vrot)$ 
assuming $\Vrot = \sqrt{2} \sigma$ for an isothermal profile. 
Finally $\phi(\Vrot)$ for ALFALFA galaxies was constructed by smoothly interpolating 
the HIVF for late-type galaxies at low velocities to  $\phi^{\text{early}}(\Vrot)$ at large 
velocities assuming a modified Schechter function. These results suggest  
that although $\Vrothi$ is not the same as $\Vrot$ for early type galaxies, a systematic correction
to $\Vrothi$ can be added to bring it closer to the '\emph{true}' rotational velocity.

We end this section with a final observation. 
We find that the HIVF and HIWF for red and blue galaxies are well separated beyond the knee of the 
velocity functions. The red population dominates the velocity function at larger velocities.
The red population also dominates the HIMF at larger masses but the difference between the red and blue 
mass functions is much smaller compared to the velocity functions. 
The blue population dominates the abundances at lower masses and velocities, 
and the red population has a slope which is 
much shallower compared to the blue population at this end.

\section{Scaling Relations for HI Selected Galaxies}
\label{sec_scaling}
In the last section we have observationally derived three distributions which describe the abundances 
of HI-selected galaxies. These are the HIMF, HIWF and HIVF. These abundances have been estimated 
based on their HI signal and the ALFALFA selection function. 
The next step is to relate HI properties to either galaxies or halos. In D20 and D21 we have presented 
the conditional HI mass functions, conditioned on optical color and/or magnitude for 
HI-selected galaxies. In what follows we will build a model to populate halos with HI. 

The Halo Abundance Matching (HAM) technique is a powerful and elegant method to match 
halo properties to galaxy properties 
\citep{2004MNRAS.353..189V,2009ApJ...696..620C,2010ApJ...717..379B,2013MNRAS.428.3121M}.
It assumes that every dark matter halo above a certain mass threshold hosts one galaxy.
In its simplest form it assumes that the most massive (or luminous) galaxy is hosted 
in the most massive halo, or a monotonic relation between galaxy mass and halo mass. 
To derive the stellar-halo mass relation of the galaxy one matches the spatial abundances 
of galaxies to that of halos.
\beq
n( > \mstar) = n (> \mhalo)
\eeq
Here  $n$ is the number density of objects above a certain mass threshold, 
$\mstar$(galaxy stellar mass) or $\mhalo$ (halo mass). To further obtain the $\mstar$-$\mhalo$  
relation one needs the stellar mass function (SMF) of galaxies (which are observationally constrained) 
and the halo mass function (which can be obtained from $N-$body simulations). 
The relation can include scatter between $\mstar$ and $\mhalo$, 
and has been shown to reproduce the clustering of galaxies  
\citep[see][and references therein]{2010ApJ...717..379B}. HAM techniques provide 
a starting point for  parametrizing  the $\mstar$-$\mhalo$ scaling relation.
It has been shown both 
observationally \citep{2013ApJ...770...57B} and with hydrodynamical 
simulations \citep{2016MNRAS.460.3100C}, that $M_{\text{peak}}$ 
(the maximum virial mass of the halo, or subhalo, over its accretion history) or $V_{\text{peak}}$ 
(the maximum $V_{\text{max}}$ -- the maximum circular velocity -- of the halo, or subhalo,  
over its accretion history) is more tightly correlated to $\mstar$ as compared to $\mhalo$.
Abundance matching $\mstar$ to either $M_{\text{peak}}$ (or $V_{\text{peak}}$) better reproduces
observed clustering data in mock catalogs. However we will show in a forthcoming paper, 
using the $\mhi-\mhalo$ relation (as opposed to $\mhi-M_{\text{peak}}$) better reproduces 
HI clustering. This relation can be refined further and 
include redshift evolution and uncertainties so that it reproduces observed 
galaxy abundances and clustering for different galaxy types 
\citep{2013ApJ...770...57B,2019MNRAS.488.3143B} across cosmic time. 

\begin{figure*}
\centering
  \begin{tabular}{cc}
    \includegraphics[width=3.4in]{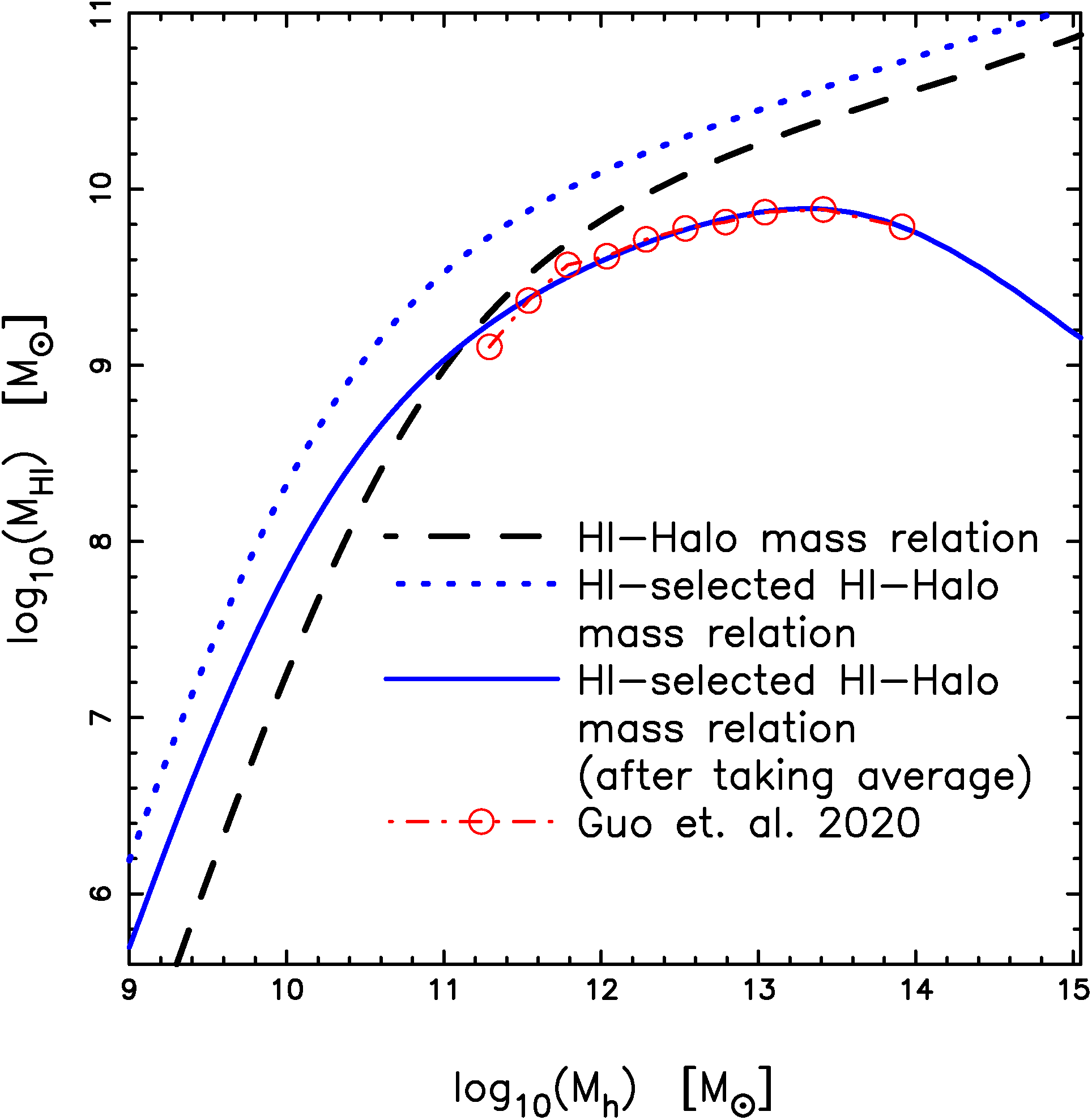} 
    \includegraphics[width=3.4in]{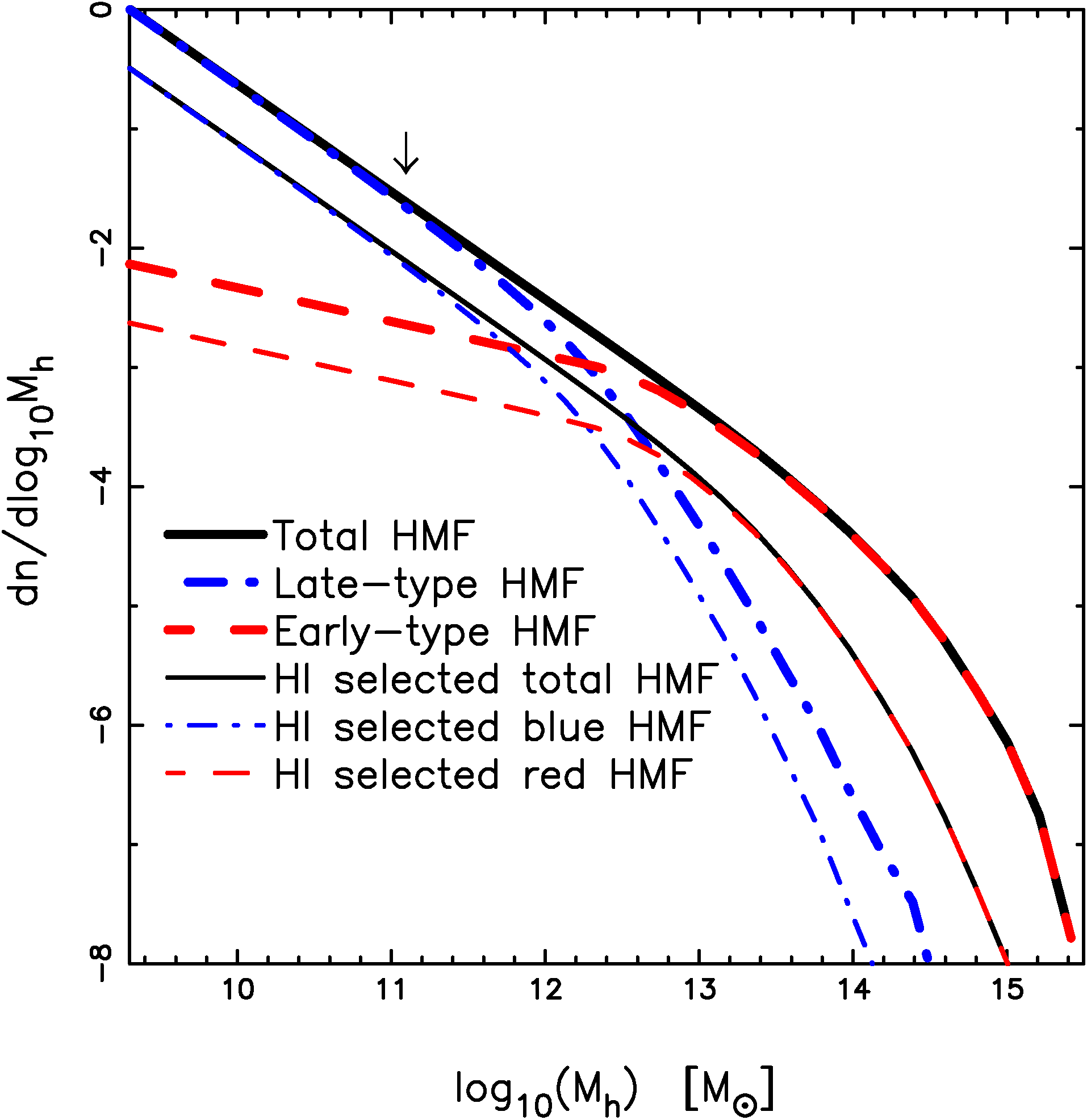}\\
  \end{tabular}
  \caption{Left panel: Average HI mass - halo mass relation 
    from \citet{2020ApJ...894...92G} is shown as open circles.
    The  dashed line is the $\mhi-\mhalo$ scaling relation 
    obtained by  abundance matching our HIMF to the HMF.
    The dotted line is the $\mhi-\mhalo$  relation 
    obtained by  abundance matching our HIMF to the HI-selected HMF 
    (see equations~\ref{eq_phihi},\ref{eq_hiscaling} and discussion).
    The solid line is the mean $\langle \mhi \rangle$ -- $\mhalo$ relation that is 
    obtained by averaging the dotted line over all the halos at any given 
    mass $[\mhalo,\mhalo +d\mhalo]$. 
    The HI-selected HMF has been defined so that the solid line matches the observed points 
    of \citet{2020ApJ...894...92G}.
    Right panel: The HMF (HI-selected HMF) is shown as the solid thick (thin) line. 
    The arrow at  $\mhalo \simeq 11$ is the resolution of the simulation catalog. 
    For masses below that we have linearly extrapolated.
    The thick dashed (dot-dashed) line represents the HMF hosting early (late) type galaxies. (See 
    figure~\ref{fig_age} and corresponding text in section~\ref{sec_scaling} 
    for the definition of early (late) type galaxies.  
    The corresponding HI-selected HMF are shown in thin lines.}
  \label{fig_mhi-mhalo}
\end{figure*}

HAM techniques have also been used to obtain the $\mhi-\mhalo$ relation 
\citep{2011MNRAS.415.2580K,2017MNRAS.470..340P}. However the basic assumption in HAM -- 
 a monotonic relation between halo mass and the galaxy property --  may not hold if 
we consider the galaxy property to be the HI mass. Unlike the stellar mass the HI content of galaxies 
is very sensitive to its local environment and may decrease with increasing halo mass. 
In massive halos, e.g. clusters and groups, 
the virial temperature is large and feedback processes, e.g. from supermassive blackholes keep the gas
ionized. It is therefore less likely, on average, 
to have a considerable amount of HI associated with such massive 
systems. These arguments are observationally supported. We find that in the sample and survey 
volume considered here, 11\% (38\%) of detections in HI are in the red (blue) cloud (D20). 
The luminous red galaxies dominate the high mass end of the galaxy SMF and 
at $\mstar = 11.3$ the number of red galaxies is $\sim 10\times$ larger than the blue galaxies 
\citep{2012MNRAS.421..621B}. These galaxies are mostly centrals \citep{2009ApJ...707.1595D} and would 
be hosted in halos of mass $\mhalo \sim 14$ \citep{2019MNRAS.488.3143B} with virial temperatures 
$T_{\text{vir}} \sim 10^7 \text{K}$. 
It would be very rare to see large amounts of HI in such systems.
Direct and more sensitive observations targeting massive galaxies, 
the GALEX Arecibo SDSS Survey (GASS)
\citep{2013MNRAS.436...34C}, and massive ETGs, the ATLAS$^{\text{3D}}$ survey 
\citep{2012MNRAS.422.1835S}, confirm these arguments. 
They find that massive galaxies are dominated by non-detections in 
HI and the limiting HI masses (upper bound based on survey sensitivity) 
is well below the knee of the HIMF ($M_* = 9.96$). We finally note that
the high mass end of the HIMF is dominated by luminous red galaxies which   
represent a  small fraction of all luminous red galaxies. As mentioned earlier 
luminous red galaxies dominate the SMF (hence the halo mass function) 
over their blue counterparts by at least a factor of 10. 
It is not surprising,therefore, to see that the HIMF is dominated by red galaxies since 
a small fraction of gas rich red galaxies is all that is needed to boost their abundances over 
that of their blue counterparts.

We have argued that HAM cannot be applied to obtain the $\mhi$-$\mhalo$ relation 
since we do not expect a monotonic correlation between HI masses and stellar or halo masses.
In what follows, given the tight monotonic relation between stellar and halo masses, we will 
use the halo and stellar masses as  proxies for each other. Although a stellar mass selected sample
should not have a monotonic relation with HI mass, we can turn this around and ask if an HI-selected 
sample like ours has a monotonic relation with stellar mass.  This is indeed true and has been seen 
both in ALFALFA \citep[][D20]{2012ApJ...756..113H} and the HI Parkes All-Sky Survey Catalog (HICAT)
\citep{2015MNRAS.447.1610M}. One can therefore expect that gas rich galaxies will, on average, 
be found in more massive halos as compared to gas poor halos. A subsample of all halos 
should host HI in such a manner that the relation between $\mhi$ and $\mhalo$ is  monotonic.

Recently \cite{2020ApJ...894...92G} used ALFALFA with an optical group catalog from SDSS  
to estimate the mean HI mass - halo mass, $\langle \mhi \rangle$ --$\mhalo$, relation by stacking HI 
on the optically selected  group catalog. The HI was stacked for the full group and separately 
for centrals in the catalog, the difference between the two therefore represents 
the contribution of total HI mass in satellites in the group.  
Due to confusion, stacking of centrals is contaminated by nearby satellites within the group.
Therefore the $\langle \mhi^{\text{cen}} \rangle$  for centrals are upper limits, the 
$\langle \mhi^{\text{sat,tot}} \rangle$ for satellites are lower limits and the  
$\langle \mhi^{\text{group}} \rangle$ is the average stacked mass in groups. The stacking procedure 
was carried out as a function of group or halo mass. In the left panel of figure~\ref{fig_mhi-mhalo}
the result of \cite{2020ApJ...894...92G}  $\langle \mhi \rangle$ --$\mhalo$ is shown 
for centrals (open circles).

\begin{figure*}
\centering
  \begin{tabular}{cc}
    \includegraphics[width=3.4in]{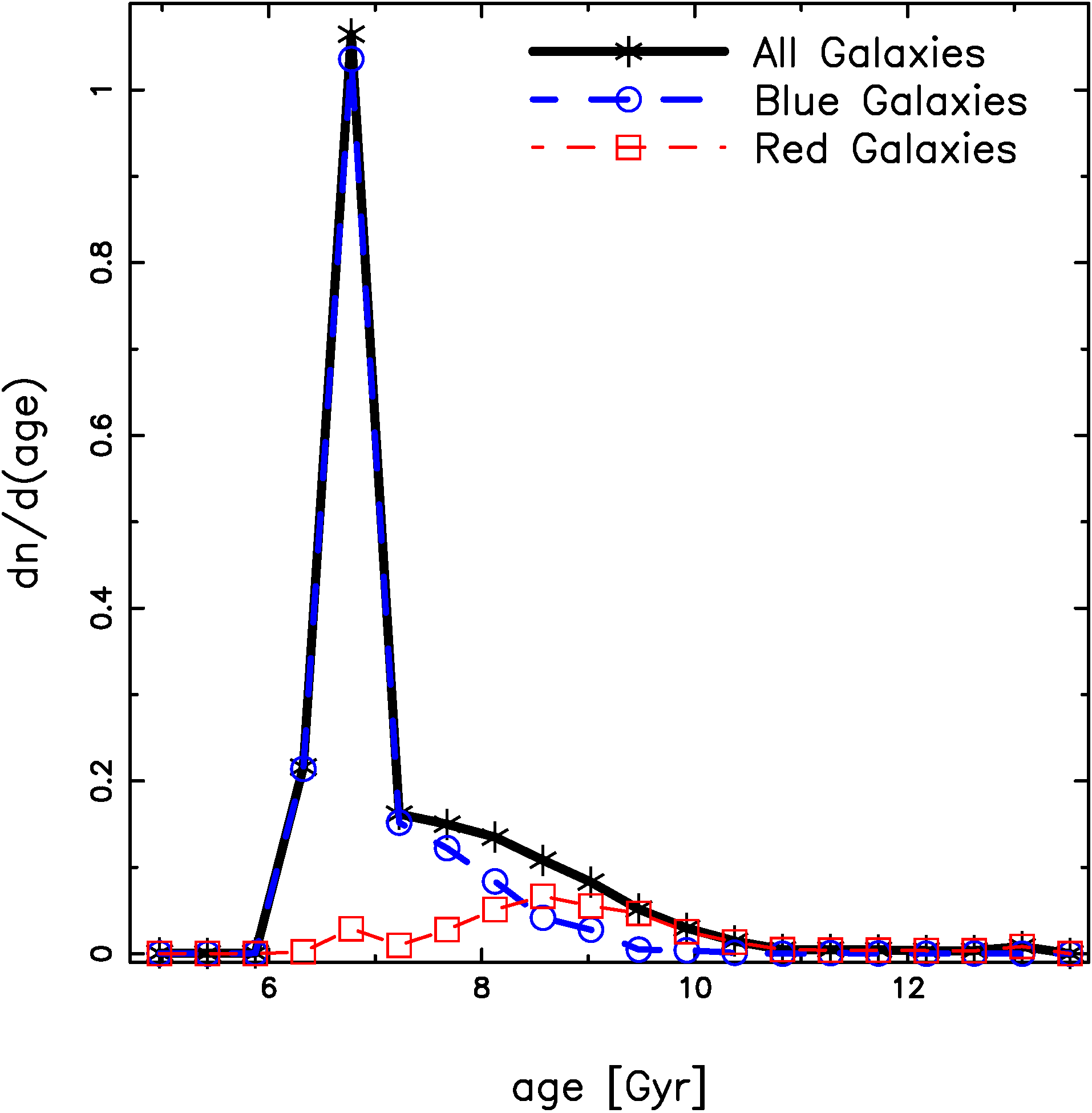} 
    \includegraphics[width=3.4in]{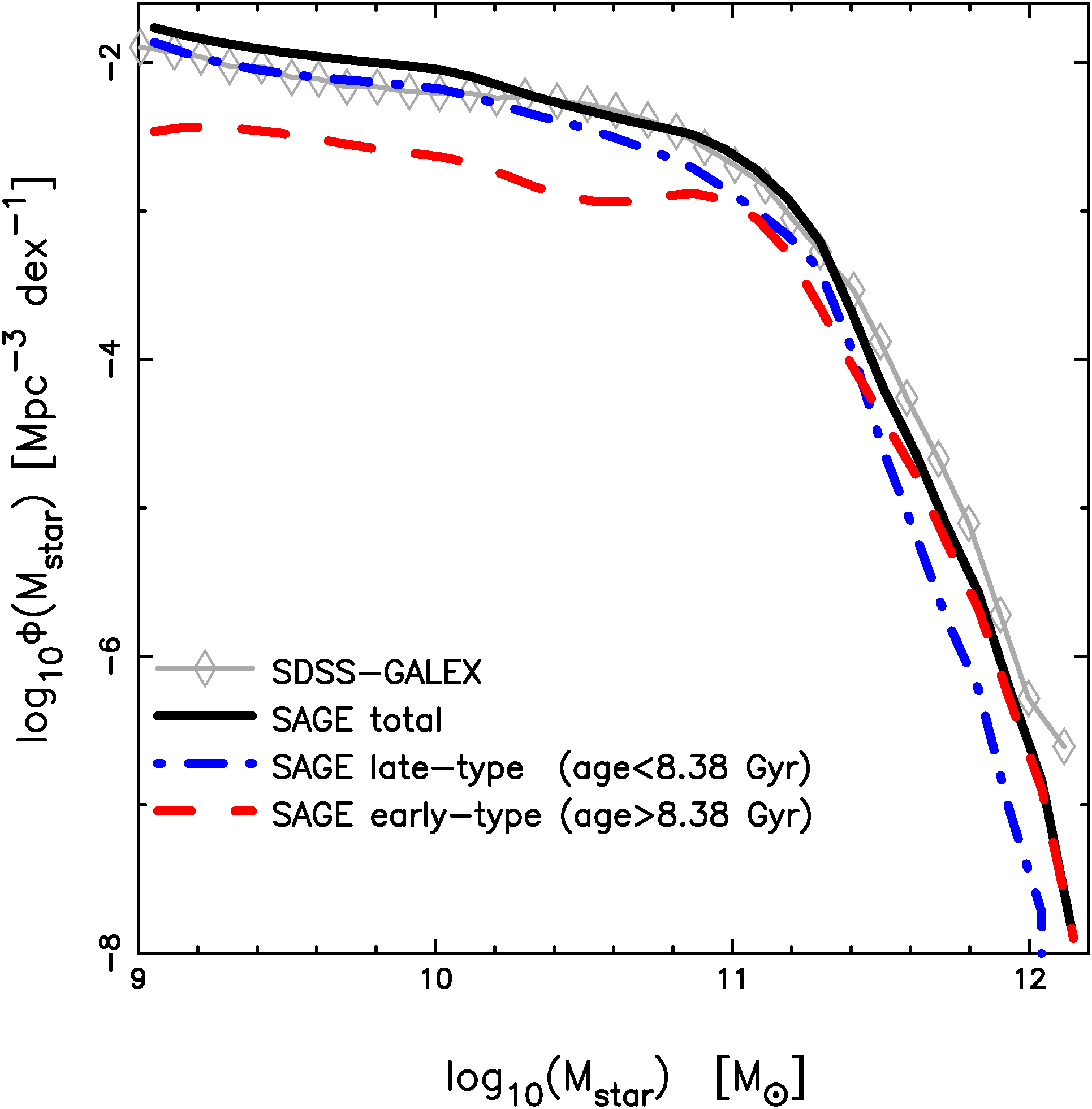}\\
  \end{tabular}
  \caption{The left panel shows the observed age (mean stellar age in Gyr) 
    distribution of the total (crosses, thick solid line)
    blue (open circles, thick dashed line)   and red (open squares, thin dashed line) sample of 
    galaxies in ALFALFA. The intersection of the age distribution of blue and red galaxies 
    at $\tage = 8.38 \text{Gyr}$ is used to classify the blue and red populations 
    as late-type and early-type galaxies respectively. 
    The right panel shows the SMF. The open-diamonds with thin solid line represents
    observational estimates of the SMF from SDSS-GALEX \citep{2013ApJ...767...50M}.
    The thick solid line is from the SAGE catalog \citep{2018MNRAS.474.5206K} which is based on the 
    MDPL2 simulation \citep{2016MNRAS.457.4340K}. The dot-dashed (dashed) line is the SMF for 
    late-type (early-type) galaxies defined as $\tage < 8.38$ ($\tage > 8.38$) Gyr 
    in the context of HI-selected galaxies (see left panel).}
  \label{fig_age}
\end{figure*}

At this stage it is important to clarify how we associate HI in galaxies and halos. 
Galaxies can be in all kinds of environments. 
They can be isolated field galaxies, central galaxies in groups
and clusters or satellite galaxies in such systems. These are distinct objects in observations. 
The halo on the other hand can be defined as a concentration of mass within a radius such that 
it is virialized inside this radius (central halo). 
Simulations show that halos have substructure and sub-substructure (and so on) 
and these objects (satellite halo) are not only self-bound but bound 
to the halo. In what follows we will refer to a halo as either a central halo or 
satellite halo and will not distinguish between them.
Most of the mass inside the radius of a halo 
is associated with the central halo (unless we have mergers of nearly equal masses),
due to which the HMF is close to the HMF for centrals. The HMF for satellites contributes
little to the HMF. 

A similar result is seen observationally for the SMF. The SMF of the centrals 
dominates the total SMF, whereas satellites contribute little to the total SMF 
\citep{2009ApJ...707.1595D}. In our definition field halos are centrals without satellites. 
We will assume that every halo (central or satellite) hosts a single galaxy. 
Similarly HI is associated with a halo via its galaxy. 
Although it is rare to find considerable amounts 
of HI in massive central galaxies of clusters, a significant fraction of HI is locked up in 
satellite galaxies as seen in observations \citep{2009MNRAS.399.1447L}. The results 
of \cite{2020ApJ...894...92G} corroborate this observation. 
As seen in figure 3 of \cite{2020ApJ...894...92G}
the total HI content in satellites can be at least 60\% in a group of mass $\mhalo \sim 13$.
However groups and clusters are less abundant as compared to lower mass objects. We therefore expect
(as in the case of the galaxy SMF) that satellites contribute little to the HIMF. This expectation
is consistent with semi-analytical models of HI \citep{2017MNRAS.465..111K}.

We use a very large publicly available dark matter 
simulation -- \emph{MultiDark Planck 2} (MDPL2) \citep{2016MNRAS.457.4340K}. 
MDPL2 is run with the publicly available code GADGET-2 \citep{2005MNRAS.364.1105S}. 
It evolves the 
matter density field, sampled by $3840^3$ dark matter particles in a comoving box of side 
$L_{\text{box}} = 1 \text{Gpc}/h$, to $z=0$. This corresponds to a dark matter mass resolution 
$m_{\text{DM}} = 1.51\times 10^9 \msun/h$. The  force resolution is $\epsilon = 5 \text{kpc}/h$.
The MDPL2 was run with a flat $\Lambda$CDM cosmology with 
($\Omega_{\Lambda},\Omega_m,h,\sigma_8,n_s$) =  (0.693,0.307,0.678,0.823,0.96) consistent 
with the Planck15 results \citep{2016A&A...594A..13P}. 
The data products include halo (central+satellite) catalogs and merger trees which were 
obtained with the help of the 
ROCKSTAR \citep{2013ApJ...762..109B} and CONSISTENT TREES \citep{2013ApJ...763...18B}
codes. The MultiDark-Galaxies \citep{2018MNRAS.474.5206K} is a catalog of galaxies built on the 
data products of MDPL2 with the semi-analytical model -- Semi-Analytic Galaxy Evolution (SAGE).
The SAGE model is tuned to reproduce the SMF in the local Universe.
Galaxy colors and $r-$band luminosities are not well reproduced in SAGE.
SAGE also provides the mean age of the stellar population of the galaxy, $\tage$ which we 
use as a proxy for color. 
The stellar mass and age of the galaxy, $\mstar$ and $\tage$,
are available for our HI sample from kcorrect and SDSS 
\citep{2009ApJ...699..486C,2014ApJS..211...17A} respectively, 
so that we can choose a galaxy population based on $\tage$.

\begin{figure*}
\centering
  \begin{tabular}{ccc}
    \includegraphics[width=2.3in]{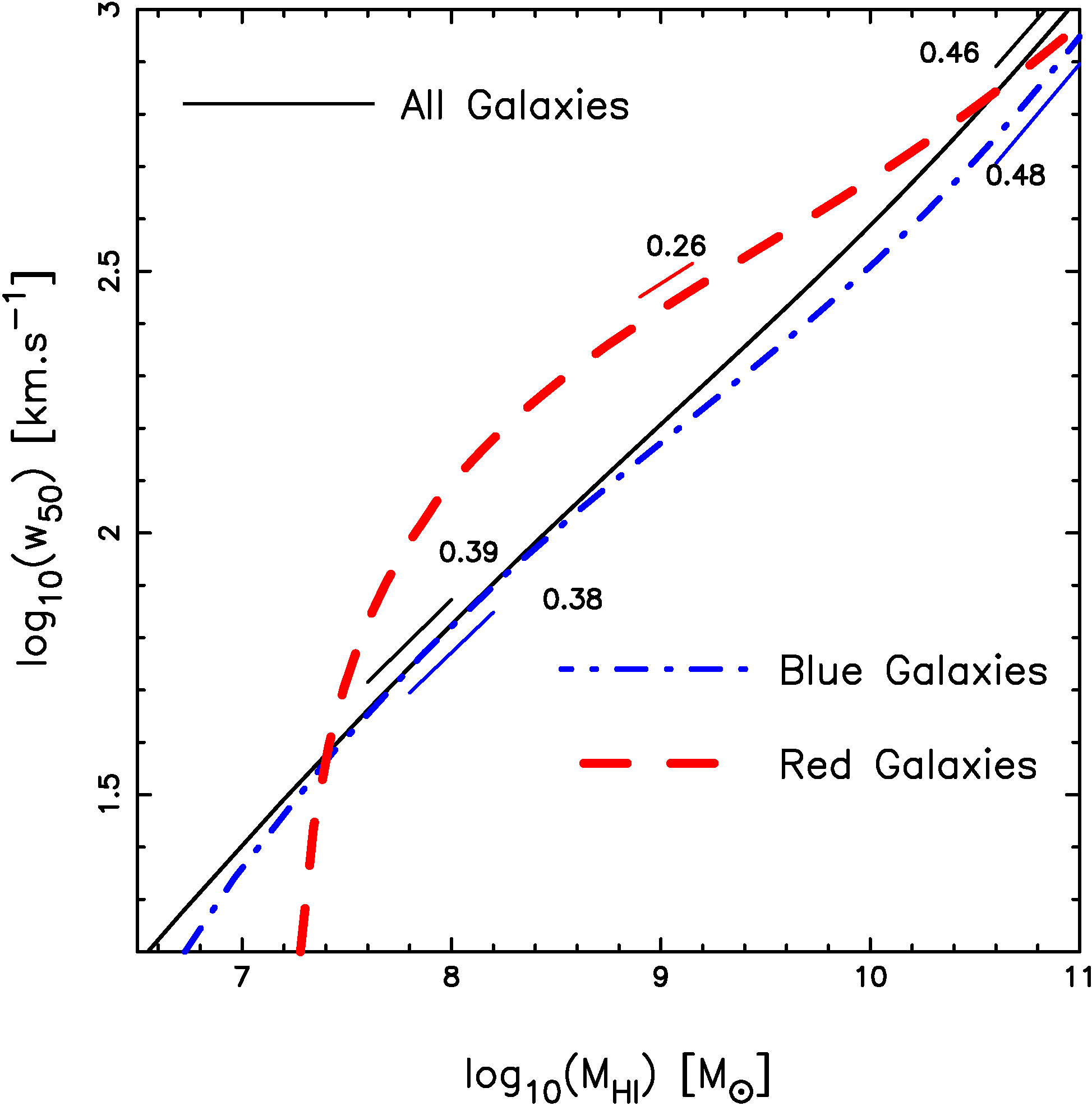} 
    \includegraphics[width=2.3in]{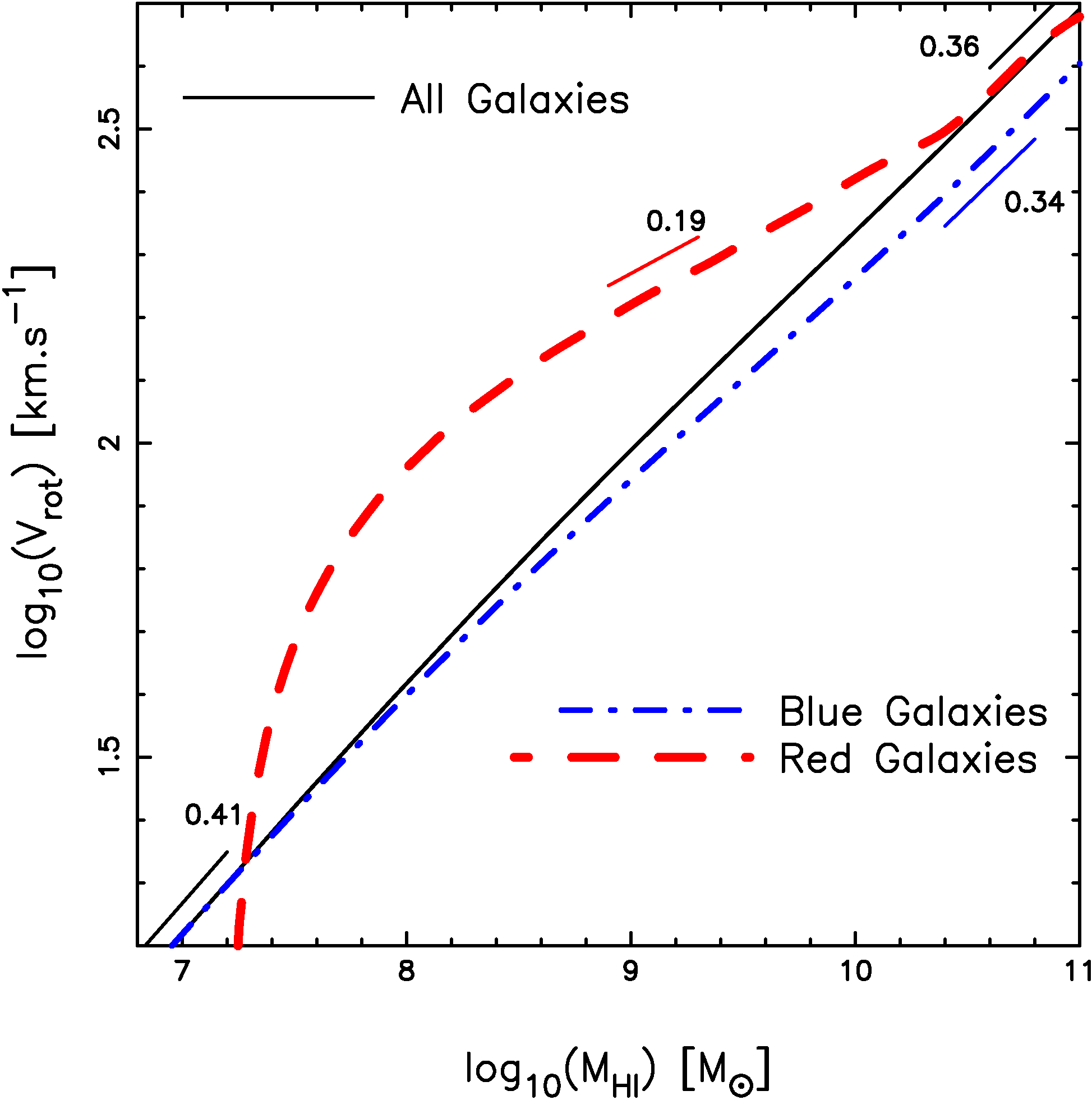}
    \includegraphics[width=2.3in]{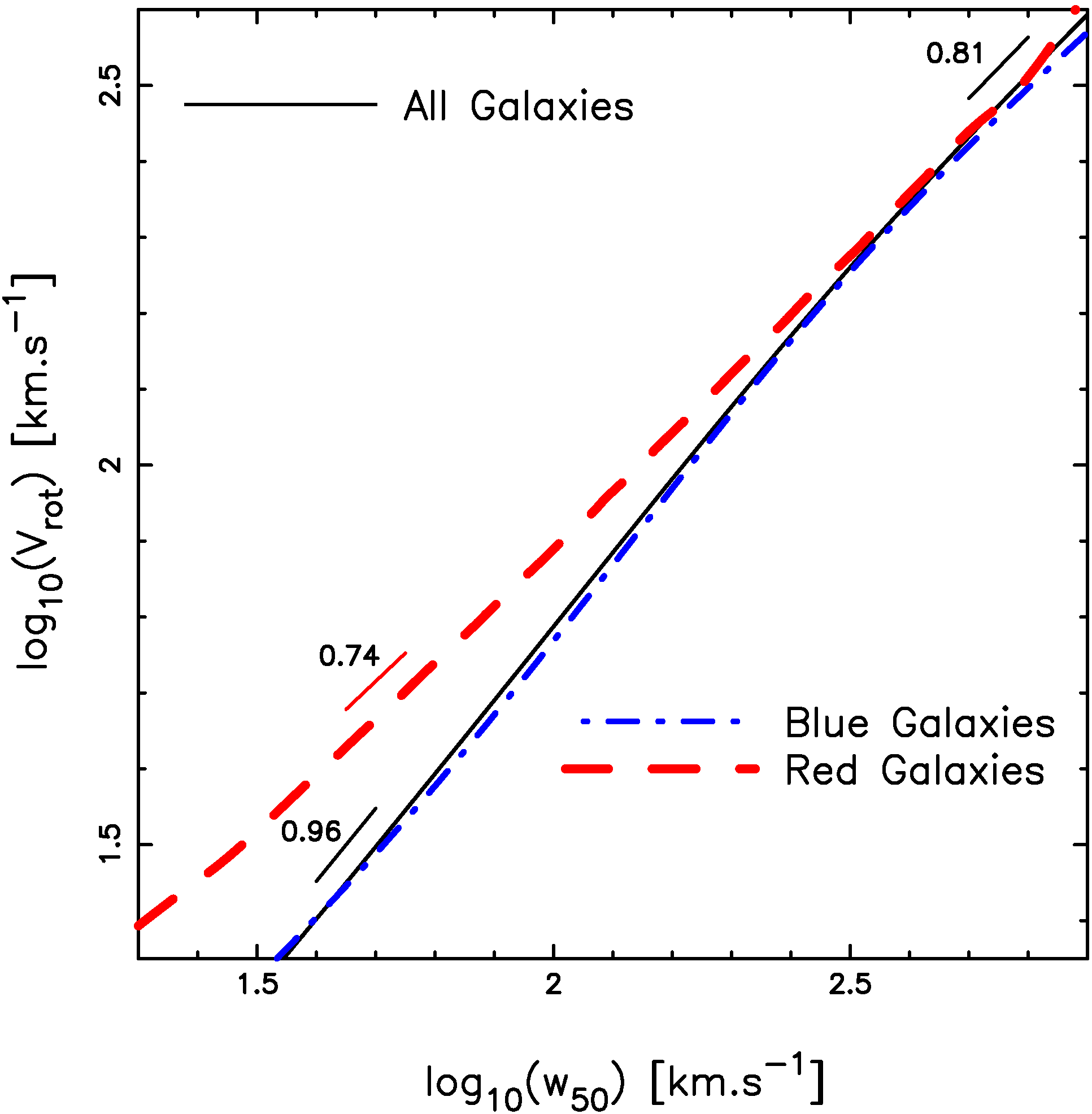}\\
  \end{tabular}
  \caption{The figure shows the scaling relations by  abundance matching the three 
    HI distributions, namely the HIMF, HIWF and HIVF amongst themselves. 
    The  left, middle and right panels show the scaling relation $\mhi-\w50$, $\mhi-\Vrothi$ 
    and $\w50-\Vrothi$ respectively for the total (solid), red(dashed) and blue (dot-dashed) samples.
  }
  \label{fig_scaling-hi}
\end{figure*}

We use the MDPL2 halo catalog to obtain the HMF. 
This is shown as the thick solid line in the right panel of figure~\ref{fig_mhi-mhalo}.
The downward arrow at $\mhalo \sim 11$ is the mass resolution of the MDPL2 catalog.
We have extrapolated the power-law to obtain the HMF at lower masses.   
We will next define the HI-selected halo mass function, $\phihi(\mhalo)$, which reproduces 
the observations of \cite{2020ApJ...894...92G} 
(open circles in the left panel of figure~\ref{fig_mhi-mhalo}) by defining it with the parametric form:
\beq
\phihi(\mhalo) = \phi(\mhalo) \times \frac{f}{1 + \left(\frac{\mhalo}{\mhalostar}\right)^\gamma}
\label{eq_phihi}
\eeq
where $\phi(\mhalo)$ is the halo mass function. $f$ is an overall fraction that reduces 
the abundances $\phihi(\mhalo)$ with respect to $\phi(\mhalo)$ and the abundance is further reduced
by $\left(\frac{\mhalo}{\mhalostar}\right)^\gamma$ for $\mhalo \gsim \mhalostar$. This is a  
three-parameter functional form that is justified by the various HI observations described earlier. 
However we point out that it is by no means unique. One can further suppress the HI mass 
at lower halo masses \citep{2010MNRAS.407..567B,2017MNRAS.469.2323P} 
since these low mass halos would host negligible amounts of HI due to the ionizing background.
However the suppression is expected to happen at circular velocities smaller than 30 km/s, 
which correspond to the smallest HI detections in ALFALFA ($\mhi \sim 7.0 - 7.5$) and is naturally 
taken care of by the ALFALFA selection function.  

We finally need to fix $\phihi(\mhalo)$ (eq.~\ref{eq_phihi}) described
by the three parameters $\left\{f, \mhalostar, \gamma \right\}$ so as to match the 
mean observed scaling relation $\langle \mhi \rangle$-$\mhalo$ for centrals 
of \cite{2020ApJ...894...92G}. As discussed 
earlier we will not distinguish between centrals and satellites while abundance matching.
HI is assigned based on halo mass and does not depend whether the halo is a central or satellite.
\cite{2020ApJ...894...92G} do not have a corresponding estimate of 
$\langle \mhi^{\text{sat}} \rangle$-$\mhalo$, but rather have $\langle \mhi^{\text{sat,tot}} \rangle$-$\mhalo$,
where  $\langle \mhi^{\text{sat,tot}} \rangle$ is the mean total HI mass in satellites
hosted in centrals of mass $\mhalo$. 
The parameters  $\left\{ f, \mhalostar, \gamma \right\}$ are fixed by $\chi^2$ minimization, so that
the scaling relation between $M_{\text{HI}}$ and $M_{\text{h}}$  at fixed halo mass
(obtained by abundance matching $\phi(\mhi)$ to $\phihi(\mhalo)$), 
when averaged over \emph{all} 
halos in the mass range $[\mhalo, \mhalo+d\mhalo]$ (given by the total HMF, $\phi(\mhalo)$)
reproduces the observed points of \citet{2020ApJ...894...92G} (open circles in the 
left panel of figure~\ref{fig_mhi-mhalo}).
The choice of $\left\{ f=0.320, \mhalostar=13.661, \gamma=0.996 \right\}$,  
obtained by this minimization procedure, results in the solid line  
in the left panel of figure~\ref{fig_mhi-mhalo}.
This constrains the HI-selected HMF, $\phihi(\mhalo)$, 
and is shown as the thin solid line in the right panel 
of figure~\ref{fig_mhi-mhalo}.

\begin{figure*}
\centering
  \begin{tabular}{ccc}
    \includegraphics[width=2.3in]{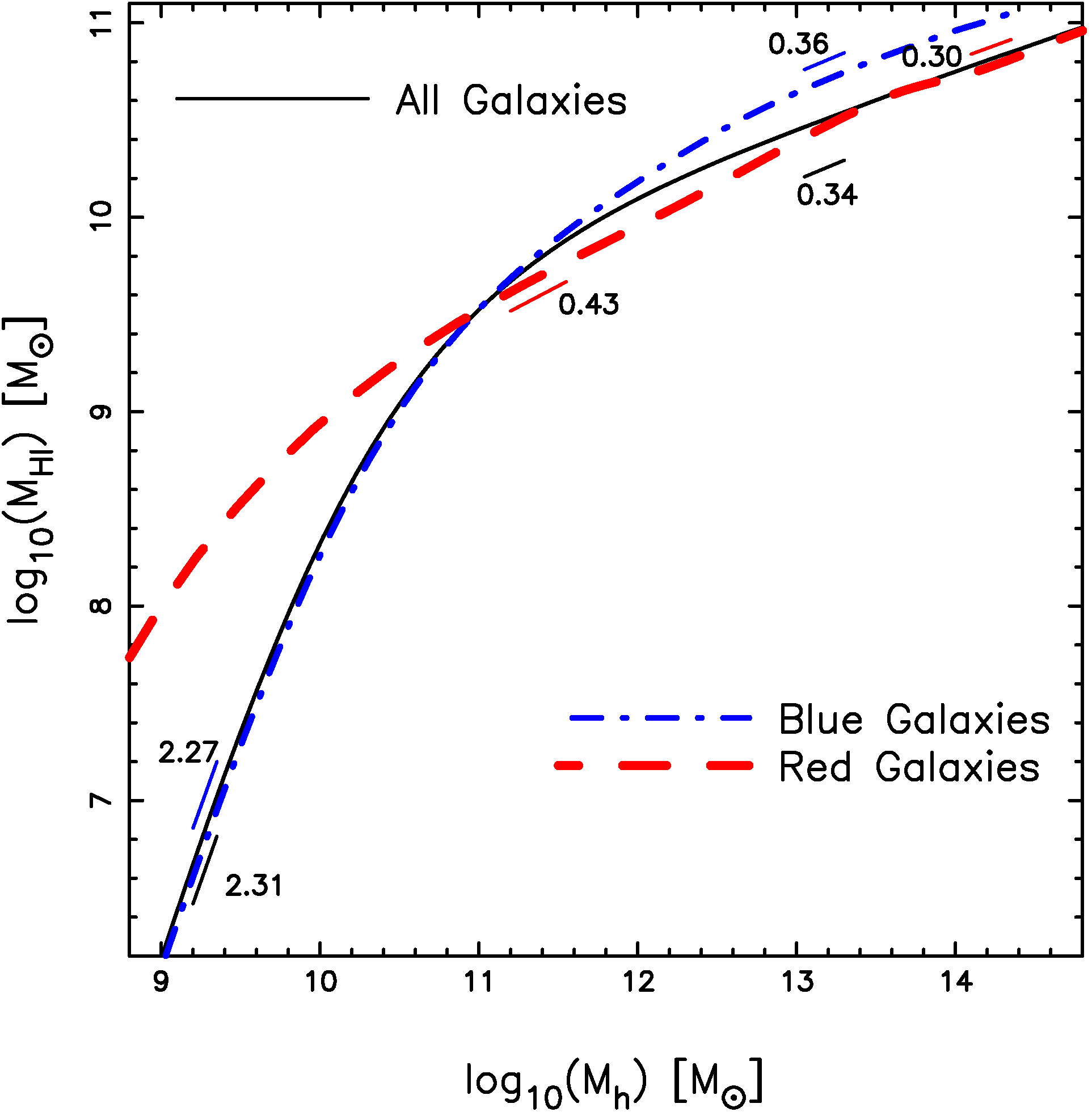}    
    \includegraphics[width=2.3in]{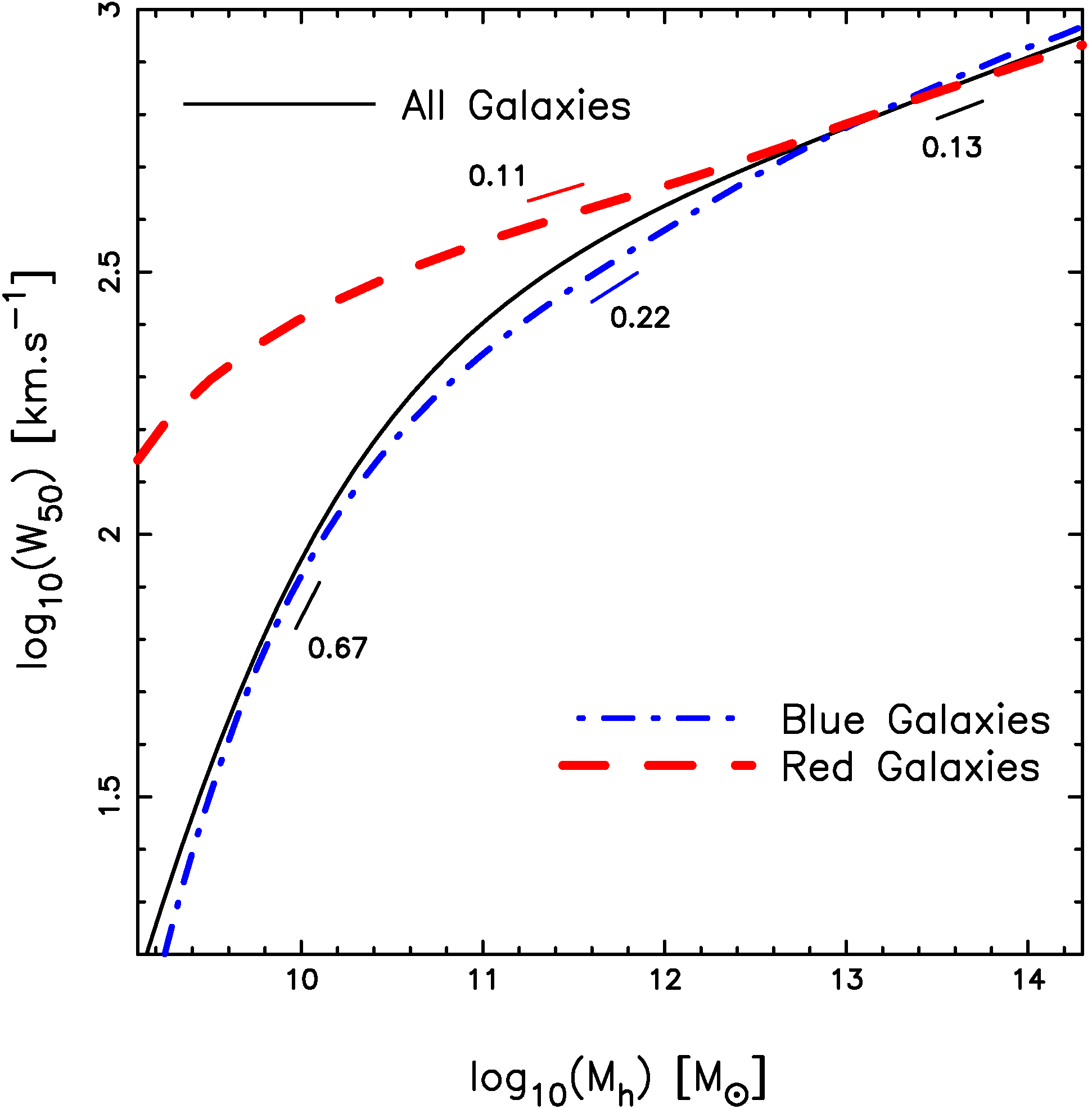}
    \includegraphics[width=2.3in]{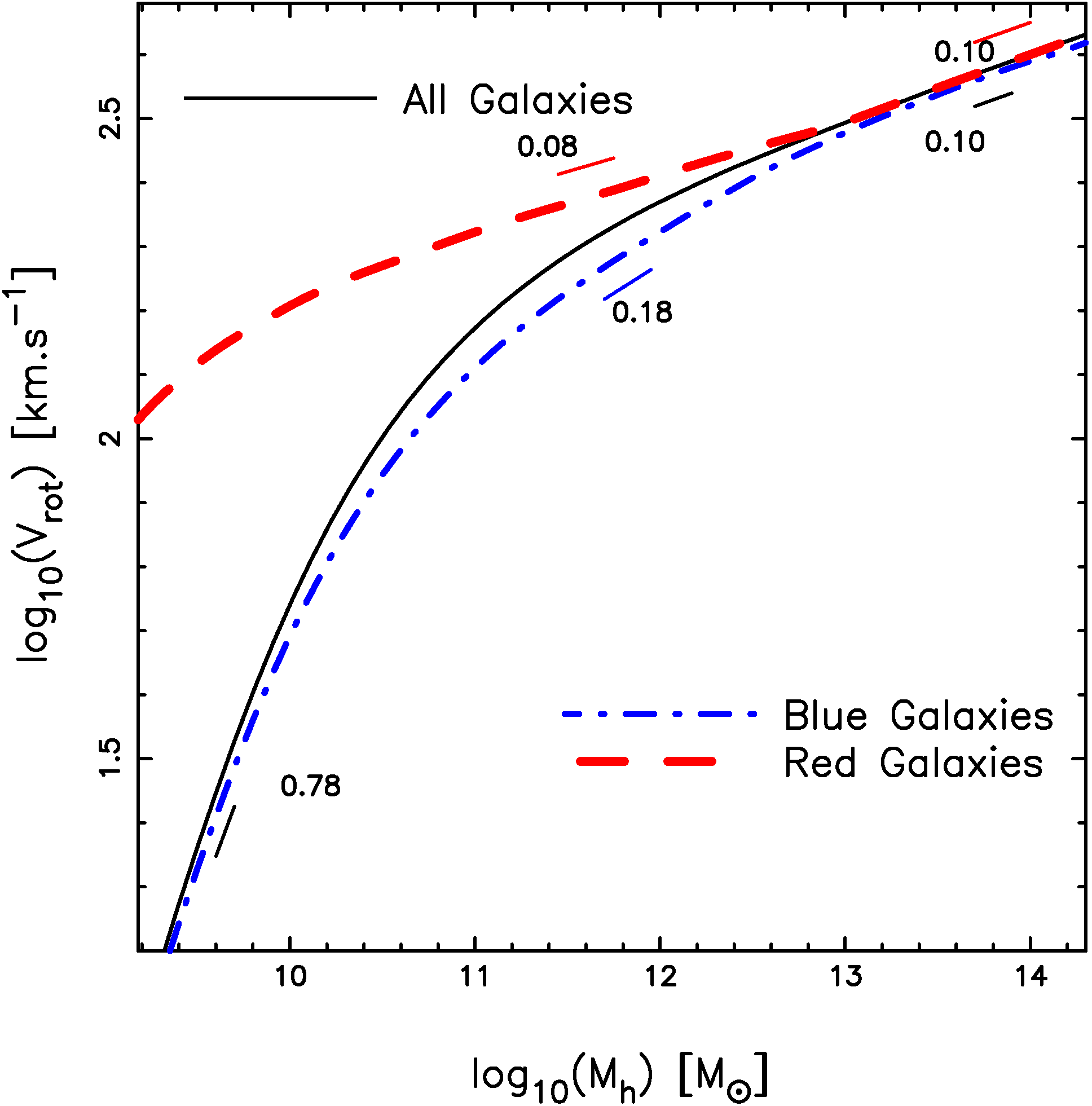}\\
  \end{tabular}
  \caption{The figure shows the scaling relations by  abundance matching the three 
    HI distributions, namely the HIMF, HIWF and HIVF with the HI-selected HMF 
    (equation~\ref{eq_phihi}). 
    The  left, middle and right panels show the scaling relation $\mhalo-\mhi$, 
    $\mhalo-\w50$ and $\mhalo-\Vrothi$ respectively 
    for the total (solid), red(dashed) and blue (dot-dashed) samples.}
  \label{fig_scaling_halo-hi}
\end{figure*}

If we abundance match the HIMF to the HI-selected HMF we obtain a scaling relation (dotted line) 
in the left panel of figure~\ref{fig_mhi-mhalo}. The dashed line is obtained by abundance 
matching the HIMF to the HMF. Since $\phihi(\mhalo)$ represents a subsample of all halos (described
by $\phi(\mhalo)$), HI
is now distributed in a smaller number of halos, thereby increasing the HI mass at fixed halo mass.
This results in a  HI-selected scaling relation (dotted line)  above the halo mass 
selected scaling relation  (dashed line). The HI selected scaling relation (dotted line) is well 
described by a double power law
\beq
\mhi(\mhalo) = \mhi^{\mathrm{A}} \frac{\left(\frac{\mhalo}{M_{\mathrm{ht}}}\right)^{\alpha}}
{\left[1 + \left(\frac{\mhalo}{M_{\mathrm{ht}}}\right)^{\beta}\right]}
\label{eq_hiscaling}
\eeq
Here $\mhi^{\mathrm{A}}$ is the amplitude, $\alpha$ is the slope of the scaling relation 
at lower masses which gets suppressed to a slope of $\alpha - \beta$ at masses greater 
than a transition halo mass, $M_{\mathrm{ht}} \ll \mhalo$. We find  
($\mhi^{\mathrm{A}}, \alpha, \beta, M_{\mathrm{ht}}$) = (9.59, 2.10, 1.76, 10.62) describes well 
the HI selected $\mhi-\mhalo$ scaling relation in figure~\ref{fig_mhi-mhalo}.

Since we are working with the red and blue populations amongst the HI-selected galaxies 
we would also like to obtain corresponding scaling relations for these populations as well.
This is where the SAGE catalog becomes useful. However, the SAGE catalog does not have accurate 
estimates of colors and magnitudes that are needed to determine the red and blue populations. 
We therefore need to come up with an approximate proxy for the red and blue populations. 
Apart from rest-frame magnitudes, kcorrect also provides  estimates of the stellar mass
and the mean age of the stellar population, $\tage$, of the galaxy 
is obtained from the Granada FSPS models \citep{2009ApJ...699..486C,2014ApJS..211...17A} from SDSS.
In the left panel of figure~\ref{fig_age} we show the age distribution of the red and blue samples in 
ALFALFA. The age distribution of the blue population has a pronounced 
peak at $\tage \sim 6.8 \text{Gyr}$ 
and drops rapidly beyond $\tage > 7.2 \text{Gyr}$. The age distribution of the red population 
has a peak at $\tage \sim 8.5 \text{Gyr}$ and the distribution is broad. Although, bimodal, 
the distribution is suppressed for red galaxies since ALFALFA primarily samples the blue cloud. 
In spite of this we can see that the red population in ALFALFA is an older, early-type population
compared to the blue population. The distributions intersect  at $\tage = 8.38 \text{Gyr}$.
We can therefore use $\tage$, as a rough proxy  for colors, with $\tage > 8.38 \text{Gyr}$ 
($\tage < 8.38 \text{Gyr}$) representing the red (blue) populations. This definition has been made 
on the basis of the ALFALFA (HI-selected) sample. A similar definition could be made on the basis 
of a stellar mass selected sample, from SDSS. However we wish to use the HI-selected HMF to abundance
match to the HI distributions therefore we will stick with this definition. 

We use this criterion to identify galaxies in SAGE as early (red) or late (blue) type galaxies.
The SAGE SMF (solid line) is plotted in the right panel of figure~\ref{fig_age}. It compares well 
with the observed SMF (open diamonds, thin line) from SDSS-GALEX \citep{2013ApJ...767...50M}. 
The dashed (dot-dashed) line is the contribution from red, early-type (blue, late-type) galaxies 
to the SMF from SAGE. The red population dominates the high mass end of the SMF whereas the blue 
population dominates the SMF at lower masses. The bimodality is however not as distinct since the 
classification was done based on an HI-selected ALFALFA sample. If it were done on a stellar
mass selected sample a clear bimodality is seen \citep{2012MNRAS.421..621B} 

In the right panel of figure~\ref{fig_mhi-mhalo} we plot the HMF corresponding to red (early-type) 
and blue (late-type) galaxies as thick dashed and thick dot-dashed lines respectively. The 
corresponding HI-selected HMF are plotted with thin lines. One can see that the early-type galaxies
dominate the HMF, and the HI-selected HMF,  
at $\mhalo \gsim 12.2$  and the late-type galaxies dominate below this mass. 

We are now in a position to obtain scaling relations between various HI properties
by abundance matching the HIMF, HIWF and the HIVF to each other. Having defined
and constrained the 
HI-selected HMF, $\phihi(\mhalo)$, we can also abundance match the HIMF, HIWF and HIVF to 
the HI-selected HMF.

In figure~\ref{fig_scaling-hi} we show the scaling relations 
$\mhi-\w50$, $\mhi-\Vrothi$ and $\w50-\Vrothi$ in the left, middle and right panels respectively.
These relations were obtained by abundance matching the HIMF-HIWF, HIMF-HIVF and HIWF-HIVF. 
This is done for the total (solid line), red (dashed line) and blue (dot-dashed line) samples. 
The scaling relation for the total sample can be thought as a galaxy-count weighted sum of 
the scaling relations of the red and blue samples. 
Since HI is primarily sampled by the blue cloud, the scaling relation 
for the full sample is closer to the scaling relation of blue galaxies. The 
scaling relations $\mhi-\w50$ and  $\mhi-\Vrothi$ are different for the red and blue samples. 
At lower masses the HI detections are primarily in the blue cloud, we therefore see a rapid drop
in the scaling relations below $\mhi = 7.6$ for the red sample. For $\mhi \in [8.5,10]$ we find 
that at fixed HI mass the red sample has larger velocity profile widths. This suggests that 
in this range the red sample is on average hosted in larger halos because 
the profile width (or rotational velocity ) is a good proxy for the halo mass. Although we have not 
invoked the HI-selected HMF at this stage, 
we see that this explanation is consistent with figure~\ref{fig_mhi-mhalo} and 
figure~\ref{fig_scaling_halo-hi}. In the right panel of figure~\ref{fig_scaling-hi} we see 
the $\w50-\Vrothi$ scaling relation for the red sample is above that of the blue sample
at lower velocities, but they asymptote to each other at larger velocities.

In figure~\ref{fig_scaling_halo-hi} we show the scaling relations 
$\mhalo-\mhi$ (left panel), $\mhalo-\w50$ (middle panel) and $\mhalo-\Vrothi$(right panel) 
which were obtained by abundance matching the HI-selected HMF-HIMF, HI-selected HMF-HIWF 
and HI-selected HMF-HIVF respectively. This was done separately for the total (solid line),
red (dashed line) and blue (dot-dashed line) samples. 
The $\mhalo-\mhi$ scaling relation is qualitatively 
similar in shape to the $\mhalo-\mstar$ scaling relation
\citep{2010ApJ...717..379B,2019MNRAS.488.3143B} and is described by a double power-law: 
a steep power law with slope $\sim 2.1$ at lower masses transitioning to a shallower power law with 
slope $\sim 0.34$ above masses $\mhalo \gsim 10.6$ (see equation~\ref{eq_hiscaling}). 
The transition  mass $M_{\mathrm{ht}} = 10.62$
for $\mhalo-\mhi$,  is more than an order of magnitude smaller than 
the transition mass of $\mhalo \simeq 12$ 
\citep{2019MNRAS.488.3143B} for the $\mhalo-\mstar$ relation. 
This suggests that baryonic processes like 
heating and feedback in larger mass halos suppress HI gas on a shorter time scale compared to 
star-formation. A double power-law is also seen in the scaling 
relations for $\mhalo-\w50$ and $\mhalo-\Vrothi$ with the transition in slopes occurring 
at  $\mhalo \sim 11$. As in figure~\ref{fig_scaling-hi} the scaling relation of the 
total sample is close to that of the blue sample.

At the low mass end ($\mhalo \leq 11$) of the $\mhalo-\mhi$ scaling relation 
we find that at fixed halo the red sample
is richer in HI compared to their blue counterparts. The situation is reversed 
at the high mass end. At the high mass end, we can turn this around.
We find at fixed $\mhi$ the halo mass is larger for the red sample compared to the blue sample.
In the middle and right panels we see a similar trend for  $\mhalo-\w50$ and $\mhalo-\Vrothi$
at lower masses. However at larger masses the relations of the blue and red samples 
asymptote to each other, suggesting that the velocity profile is a good descriptor of 
the halo mass irrespective of galaxy type.

\begin{figure}
\centering
  \includegraphics[width=\columnwidth]{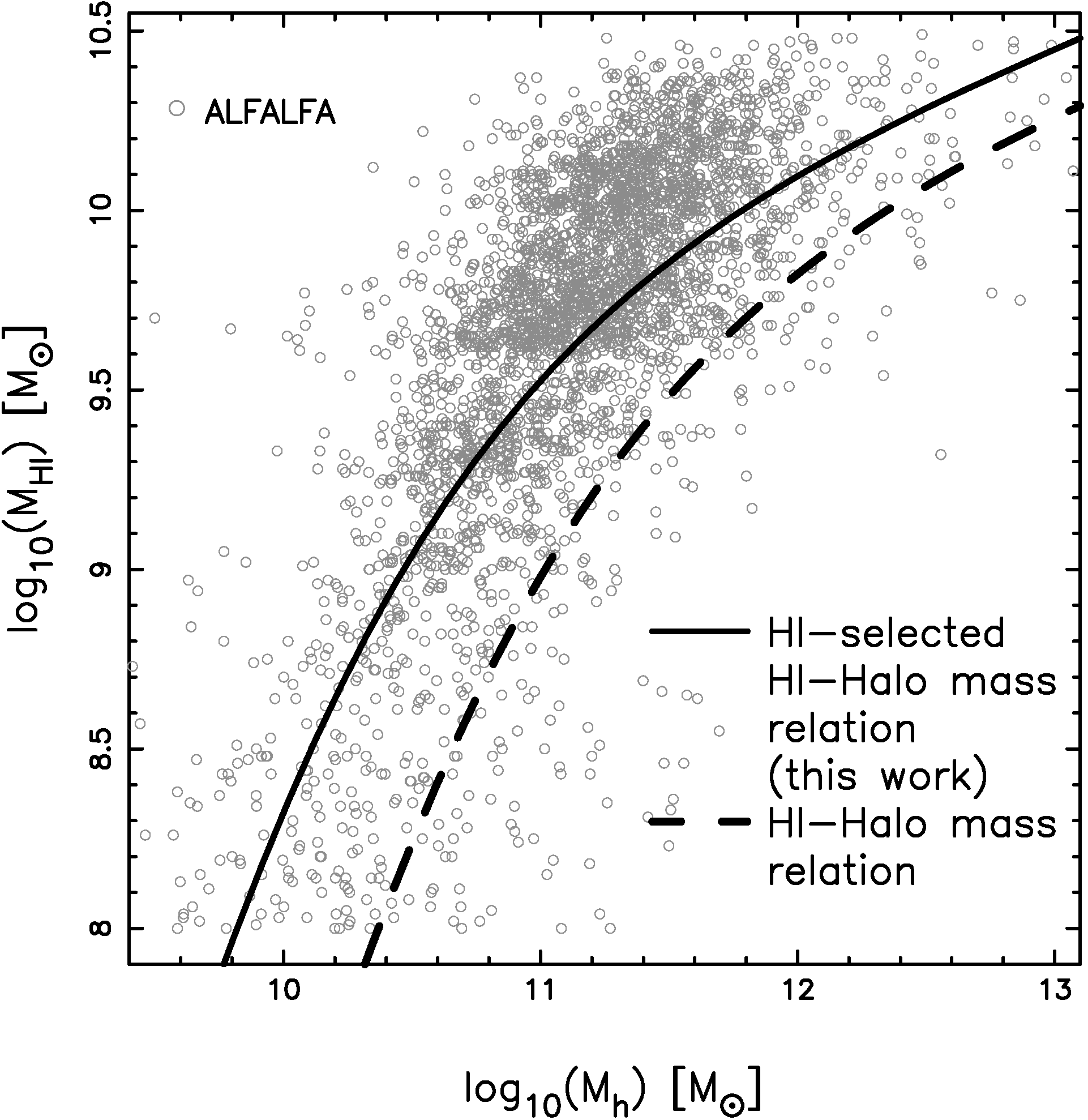}    
  \caption{The $\mhi-\mhalo$ relation (open circles) for a 
    volume-limited subsample generated from the ALFALFA survey. The $\mhalo$ masses 
    were estimated using the $\mstar-\mhalo$ relation of  
    \citet{2013MNRAS.428.3121M}.   
    The  solid (dashed) line is the $\mhi-\mhalo$ relation 
    obtained by abundance matching the HIMF with the HI-selected HMF (HMF).}
  \label{fig_scaling_compare}
\end{figure}

We end this section by comparing the $\mhi-\mhalo$ relation obtained with ALFALFA data.
This is a useful consistency check of our results with data. 
We create five volume limited samples in equal bins of mass from ALFALFA,
in the mass range $\mhi \in [8.0,10.5[$. The five volume limited samples are disjoint sets.
We combine them to create a final volume limited sample. We use the stellar masses of these galaxies
and convert them into halo masses using the tight $\mstar-\mhalo$ scaling relation of 
\citet{2013MNRAS.428.3121M}.
Using other relations, e.g. \cite{2010ApJ...717..379B,2019MNRAS.488.3143B} 
does not result in significant changes to our 
result in figure~\ref{fig_scaling_compare}. The solid line is the scaling relation that we obtain 
by abundance matching the HIMF to the HI-selected HMF (equation~\ref{eq_phihi}) and the dashed
line is the scaling relation obtained by abundance matching the HIMF to the HMF. 
Clearly the scaling that we obtain (solid line) is in better agreement with the data 
as compared to the scaling relation obtained by abundance matching the HIMF and HMF (dashed line).

\section{Discussion and Summary}
\label{sec_summary}
In this work we have used data from the ALFALFA survey to obtain 
the HIMF, HIWF and HIVF for HI-selected galaxies. The survey volume
that we have considered overlaps with SDSS and allows us to 
also look at these abundances for the red and blue population of galaxies. 
We then use recent observations from ALFALFA
which estimate $\langle \mhi \rangle$ -- $\mhalo$ relation in massive centrals 
\citep{2020ApJ...894...92G} to finally estimate an HI-selected HMF, $\phihi(\mhalo)$ 
(equation~\ref{eq_phihi}).
$\phihi(\mhalo)$ is parameterized by three parameters which are fixed 
to match the observed  $\langle \mhi \rangle$ -- $\mhalo$  relation. 
Although an upper bound, this relation explains the $\mhi-\mhalo$ relation 
for an HI-selected sample (figure~\ref{fig_scaling_compare}). 
We then use a semi-analytic galaxy catalog, SAGE, which was generated from a large  
simulation, MDPL2, to further 
obtain the HI-selected HMF for red (early-type) and blue(late-type) galaxies.

\begin{figure*}
\centering
  \begin{tabular}{cc}
    \includegraphics[width=3.4in]{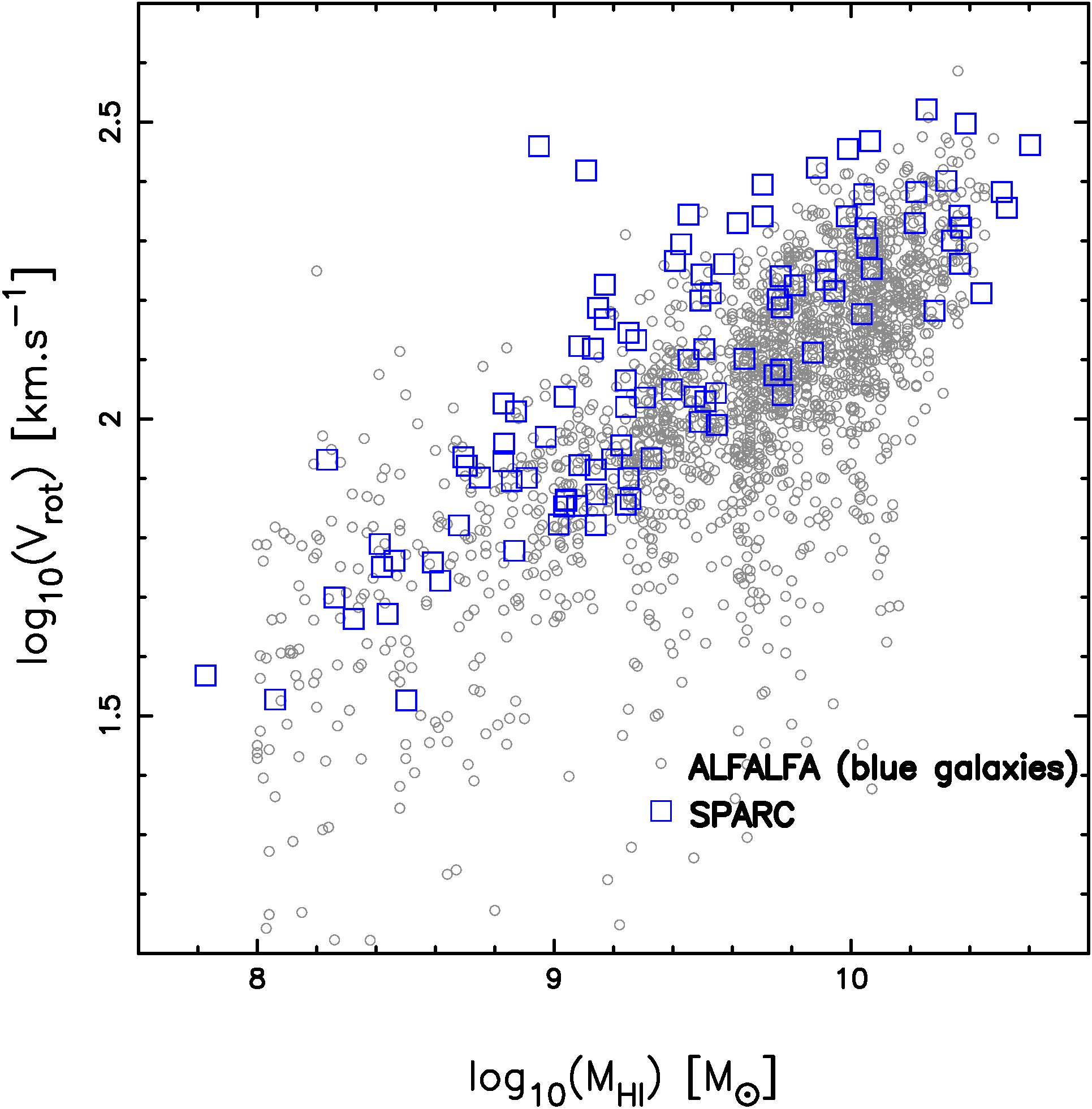}    
    \includegraphics[width=3.4in]{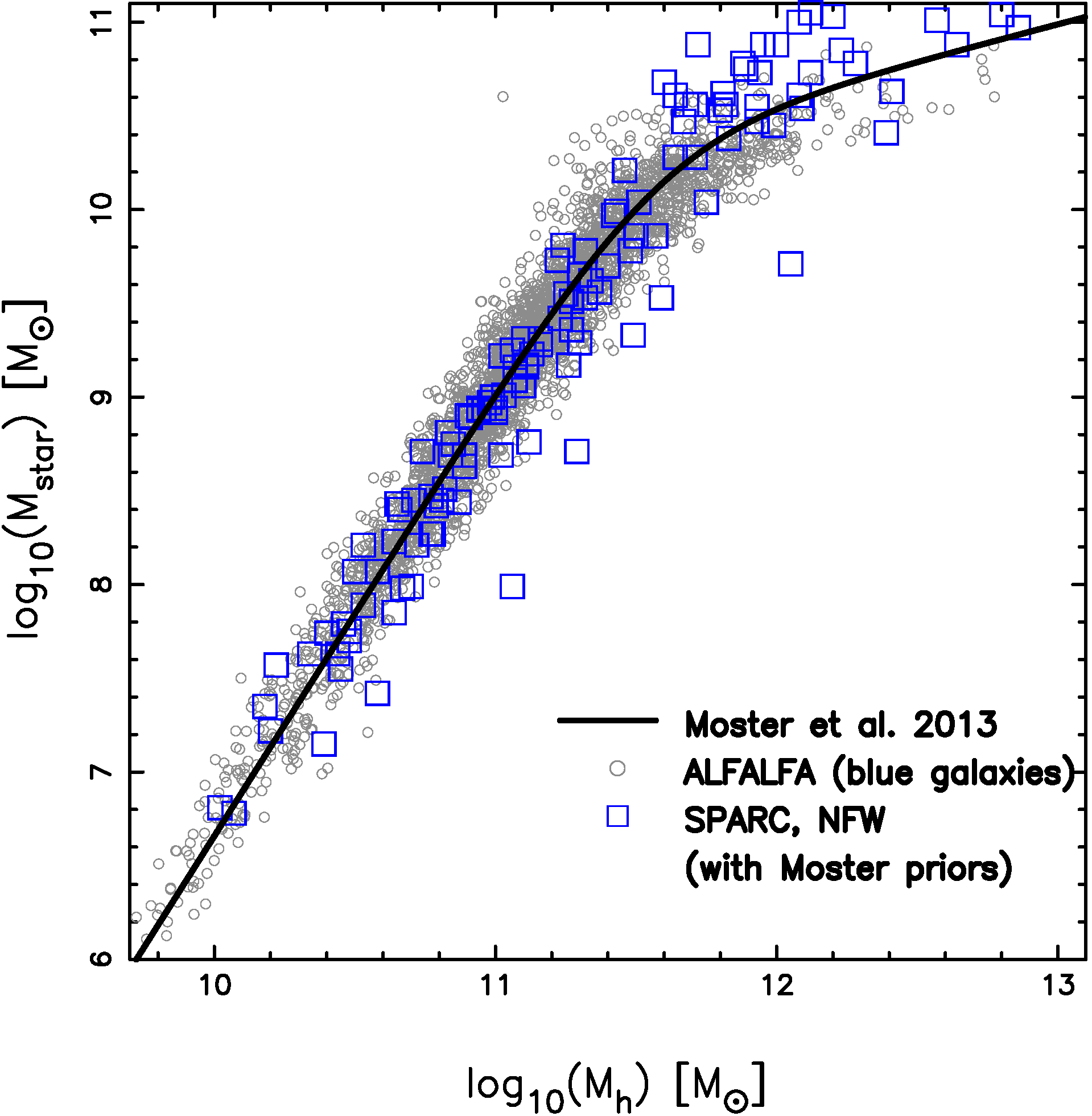}\\
  \end{tabular}
  \caption{Left Panel: A comparison of HI properties, $\mhi$-$\Vrot$, 
    for a volume limited sample of blue galaxies in ALFALFA (open circle)
    and SPARC (open square). Right panel: Same as left panel but for $\mhalo$-$\mstar$. 
    The solid line is the $\mstar$-$\mhalo$ relation of \citet{2013MNRAS.428.3121M}.}
  \label{fig_mhi-vrot}
\end{figure*}

\begin{figure*}
\centering
  \begin{tabular}{cc}
    \includegraphics[width=3.4in]{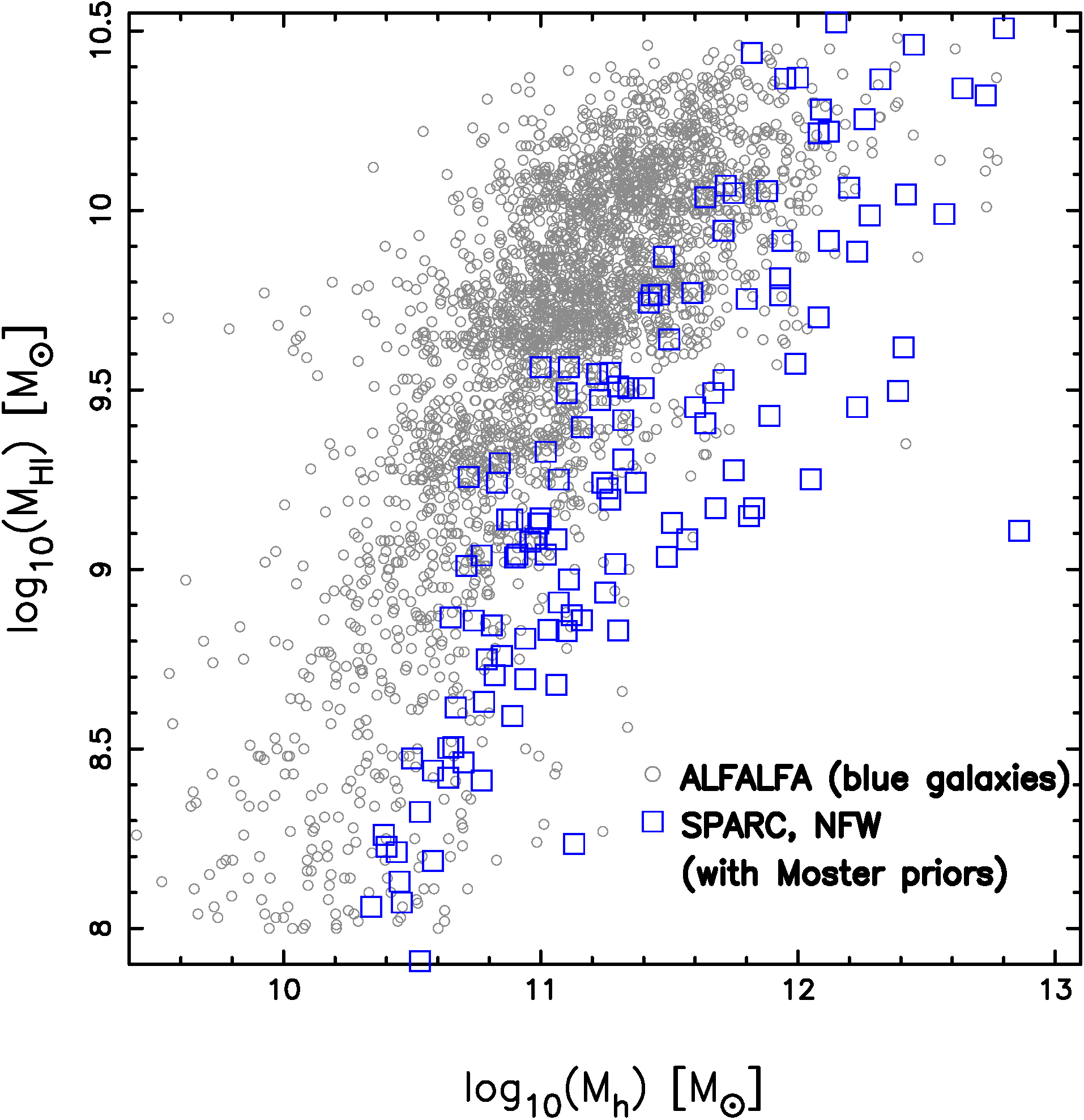}
    \includegraphics[width=3.4in]{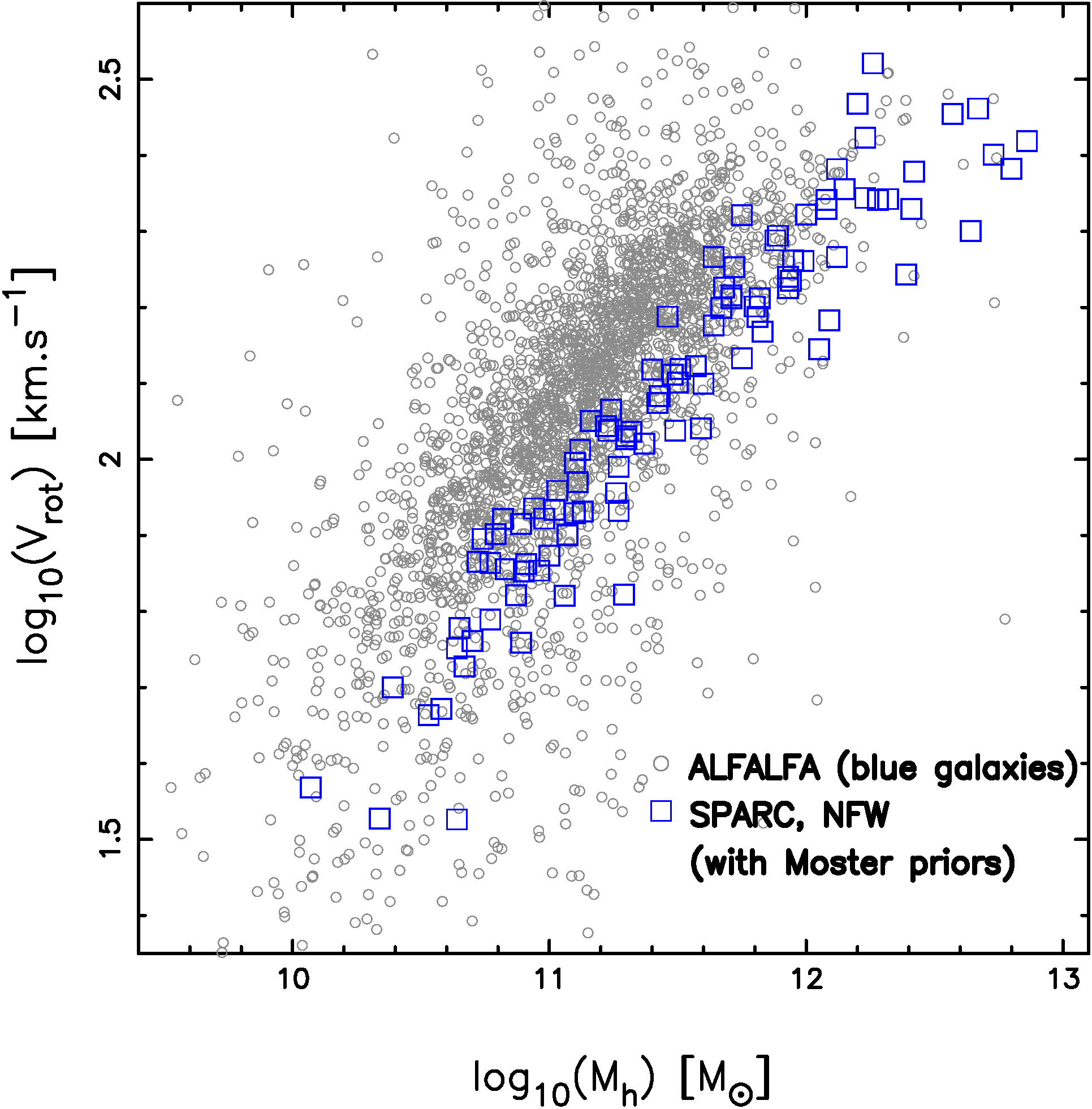}\\
  \end{tabular}
  \caption{Left Panel: A comparison of $\mhi$-$\mhalo$, 
    for a volume limited sample of blue galaxies in ALFALFA (open circle)
    and SPARC (open square). Right panel: Same as left panel but for $\Vrot$-$\mhalo$.}
  \label{fig_mhalo-mhi}
\end{figure*}

There are a number of assumptions while obtaining the HI-selected HMF for red and blue galaxies.
We have assumed that age is a proxy for color, justified observationally (figure~\ref{fig_age}). 
Although it gives a qualitatively similar bimodal 
behavior to that seen in the observed \citep{2009ApJ...707.1595D,2012MNRAS.421..621B} 
SMF (figure~\ref{fig_age}), it is by no way an exact proxy. The second assumption 
is that the stellar ages from SAGE are accurate. The stellar ages from SAGE are based 
on the halo merger trees (or growth histories) 
and the various assumptions of their model. In spite of this it gives a qualitatively 
and physically reasonable bimodal distribution for age. Finally we have used the same 
relation (equation~\ref{eq_phihi}) to obtain an HI selected halo mass function 
for both the red and blue galaxies. This may not be true. In order to distinguish 
between them we would need the stacking results of \citet{2020ApJ...894...92G} to be made 
for red and blue galaxies separately which is not available. With these assumptions in mind 
we stress that the HI-halo scaling relations for the red and blue sample 
may not be completely accurate, but should be thought of as a result which should be revisited 
once more data (both from observations and simulations) becomes available in the future. However the 
various HI scaling relations (figure~\ref{fig_scaling-hi}) are robust for the total, 
red and blue samples since there are no model assumptions. Similarly 
the HI-halo scaling relations for the total sample are also robust.

Recently  \cite{2019ApJ...886L..11L} presented the HI-selected HMF for late-type galaxies.
They used a scaling relation $\w50-\mhalo$ from 175 late-type galaxies 
in the Spitzer Photometry and Accurate Rotation Curves 
\citep[SPARC,][]{2016AJ....152..157L} catalog to determine the halo mass function of early-type 
HI-selected galaxies in HIPASS. The 2DSWML method was used to obtain the 
HI-selected HMF for early type galaxies by binning in $\mhalo$ instead of $\w50$ or $\Vrot$. 
The SPARC catalog uses near-infrared (NIR) Spitzer photometry ($3.6\mu$m) to trace the stellar mass 
distribution in galaxies. This is important to break  the star-halo degeneracy  
\citep{2016AJ....152..157L} when mass modeling galaxies. Additionally it relies on 
HI/H$\alpha$ rotation curve measurements over the past 3 decades. The rotation curves 
are finally fit assuming a halo profile, which results in a halo mass $M_{200} \equiv \mhalo$
estimate of the galaxy \citep{2019MNRAS.482.5106L,2020ApJS..247...31L}. 
The halo mass estimates of \cite{2019MNRAS.482.5106L} were made after imposing
the halo mass -- concentration relation \citep{2014MNRAS.441.3359D} and the stellar mass -- halo mass 
relation \citep{2013MNRAS.428.3121M} as priors. Imposing these priors reduces the scatter between 
the halo mass and other properties of the galaxy like $\mstar, \mhi, \Vrot, \w50$. 
We will consider halo mass estimates with these priors imposed in order to be consistent with 
\cite{2019ApJ...886L..11L}. We however point out that 
\cite{2020ApJS..247...31L} have also released halo mass estimates which relax these assumptions.
The systematic differences between SPARC and ALFALFA (figure~\ref{fig_mhalo-mhi}) which we discuss 
next,  persist nevertheless, irrespective of halo mass estimates. 
Although the halo mass estimates exist for many profiles, we will 
only consider the estimate based on the NFW \citep{1996ApJ...462..563N} profile. 
Finally 
SPARC extracts profile widths, $\w50$ of these galaxies from the \emph{Extragalactic Distance Database} 
\citep{2009AJ....138..323T,2009AJ....138.1938C}. 
The galaxy properties e.g.$\mstar, \mhalo, \mhi, \Vrot, \w50$ 
in SPARC forms a near homogeneous data set \citep{2016AJ....152..157L}.

In figure~\ref{fig_mhi-vrot} we compare the $\mhi-\Vrot$ and the $\mhalo-\mstar$ relation between 
a volume limited sample of blue galaxies in ALFALFA (open circle) and SPARC (open square).
We take the same volume limited sample as shown in figure~\ref{fig_scaling_compare} 
and choose galaxies with inclinations $i < 45^o$ to reliably obtain $\Vrot$ \citep{2010MNRAS.403.1969Z}
from $\w50$ after correcting for inclination. The volume limited sample 
is the same in figure~\ref{fig_mhi-vrot} and \ref{fig_mhalo-mhi}. However in the figures which 
involve $\Vrot$ the sample is smaller since it excludes galaxies with  $i > 45^o$. 
The scaling of HI properties (left panel) 
between  ALFALFA and SPARC agree with each other. The SPARC 
sample is also homogeneously sampling the range of   $\mhi-\Vrot$ covered by ALFALFA. 
In the right panel of  figure~\ref{fig_mhi-vrot} we compare the $\mstar$-$\mhalo$ relation
between the ALFALFA and SPARC samples. For ALFALFA we do not have an independent measure 
of $\mhalo$, we have therefore used 
the $\mstar-\mhalo$ scaling relation
from \cite{2013MNRAS.428.3121M} (solid line) 
to convert $\mstar$ into $\mhalo$ after accounting for scatter. 
The SPARC sample agrees with the \cite{2013MNRAS.428.3121M} relation and 
therefore with 
the ALFALFA points, confirming again that this relation is tight. Here too we see that 
SPARC  homogeneously samples the $\mstar$-$\mhalo$ range and the scaling is broadly consistent 
with ALFALFA. 

We now compare the $\mhi-\mhalo$ and the $\Vrot-\mhalo$ relation between ALFALFA and SPARC
in the left and right panels of figure~\ref{fig_mhalo-mhi}. We  see a marked difference
between ALFALFA and SPARC in both these figures. We point out that if we replaced 
$\mhalo$ with $\mstar$ the systematic differences remain. 
Clearly the joint distribution between an HI property ($\mhi$ or $\Vrot$) 
and optical property ($\mstar$ which is a proxy for $\mhalo$) is different between 
ALFALFA and SPARC. Although SPARC homogeneously samples individual properties, 
it is clearly biased compared to ALFALFA. For ALFALFA the selection function is well understood,
however SPARC does not have a corresponding selection function since it relies 
on individual objects. At fixed, $\mhi$ or $\Vrot$ SPARC predicts a larger halo mass. 
We would therefore expect the HI-selected HMF for late-type galaxies using the  SPARC 
scaling ($\mhalo-\Vrot$ or $\mhalo-\mhi$),
to be offset towards larger halo masses when compared to our result.

\begin{figure}
\centering
    \includegraphics[width=3.4in]{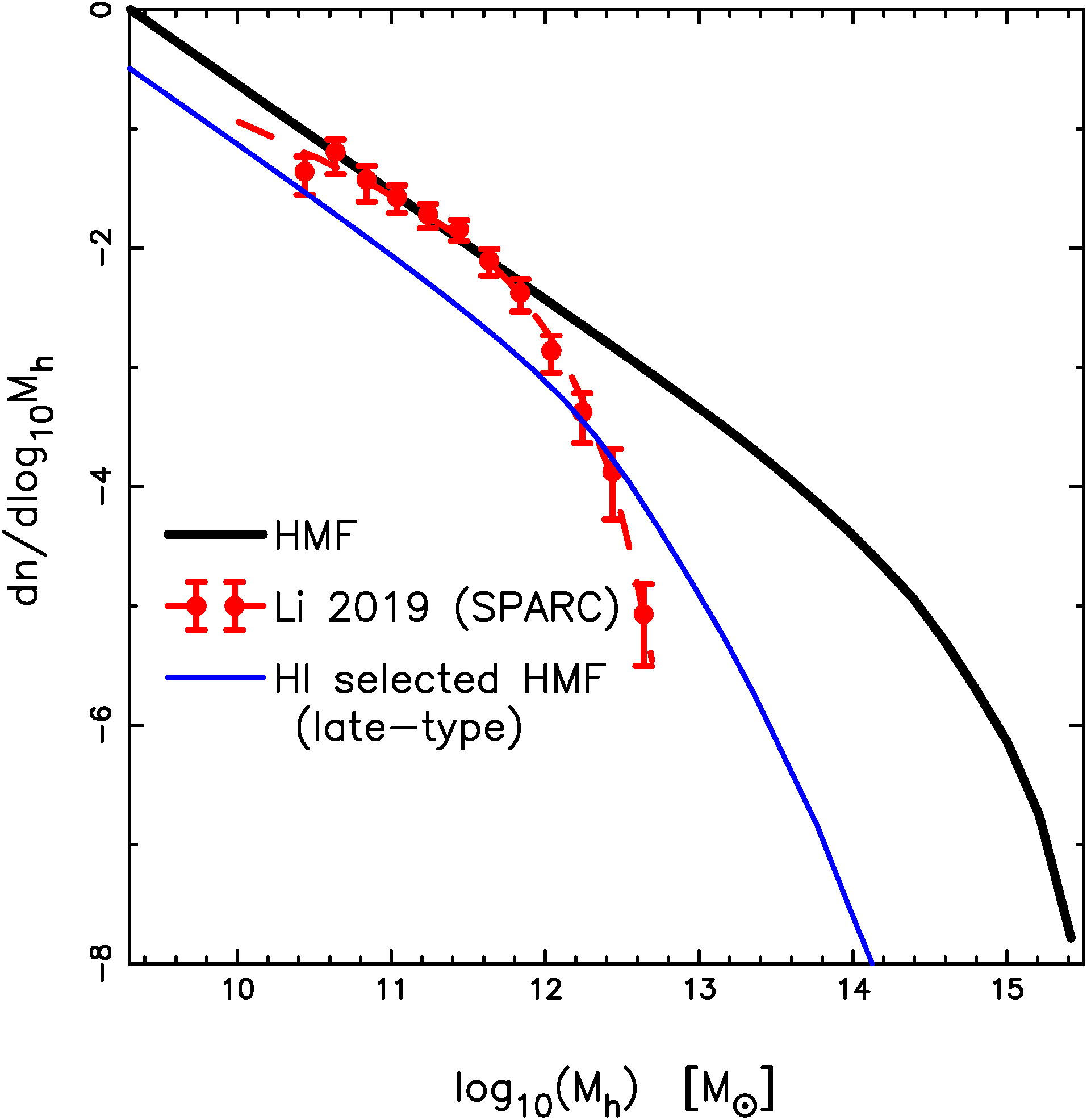}
  \caption{The thick line is the HMF. The thin line is our estimate of 
    the HI-selected HMF for late-type (or blue) galaxies from ALFALFA. 
    The dashed line is the HI-selected HMF for late-type galaxies by  
    \citet{2019ApJ...886L..11L} based on SPARC and HIPASS data.
    These halo mass estimates are based on the NFW \citep{1996ApJ...462..563N}
    profile, and assume the halo mass-concentration relation \citep{2014MNRAS.441.3359D}
    and the stellar-halo mass relation \citep{2013MNRAS.428.3121M} as priors.}
  \label{fig_hihmf}
\end{figure}

This is shown in figure~\ref{fig_hihmf}. The estimate by \cite{2019ApJ...886L..11L}
is offset by about 0.7 dex towards 
the right at lower masses and has a sharper exponential drop compared to our results (thin line).  
The offset is consistent with the offset of 0.77 dex which we estimate from the left panel of 
figure~\ref{fig_mhalo-mhi}. However the sharp drop at larger masses cannot be reconciled 
with our results by simple scaling arguments. 
We leave the investigation of such issues to future work.

We summarize our results below. 
\begin{itemize}
  \item We have shown that the HIWF and HIVF are described by modified Schechter functions. 
    Both these abundances are well separated for the red and blue populations at larger velocities. 
    The red population dominates the high velocity end and the blue population dominates the 
    velocity function at the knee and lower velocities. 
    
  \item A qualitatively similar result is seen for the HIMF. However unlike 
    the HIWF and HIVF we find that the differences in the 
    HIMF for the red and blue populations, at the high mass end, is less pronounced.

  \item Using the recent observational $\langle \mhi \rangle$--$\mhalo$ 
    relation of \cite{2020ApJ...894...92G}
    We have estimated the HI-selected HMF (equation~\ref{eq_phihi}) which represents
    the abundances of halos based on HI mass. 
    Using  semi-analytic model -- SAGE-- based galaxy catalog, 
    we have also estimated the HI-selected HMF for the red and blue galaxies. 
    
  \item Using these six abundances which describe HI rich galaxies
    we have obtained scaling relations between HI properties ($\mhi-\w50-\Vrot$)
    and HI-halo properties $\mhi-\mhalo$, $\w50-\mhalo$ and $\Vrot-\mhalo$ 
    (See figures~\ref{fig_scaling-hi} and \ref{fig_scaling_halo-hi}).

  \item The  $\mhi-\mhalo$ scaling relation is robust and consistent with a volume 
    limited sample in ALFALFA. It is described by a steep 
    power law slope $\sim 2.10$  at small masses and transitions to a shallower 
    slope $\sim 0.34$ at masses larger than $M_{\text{ht}} = 10.62$. 
    It has a shape similar to the $\mstar-\mhalo$ 
    \citep{2019MNRAS.488.3143B} but the transition halo mass, $M_{\text{ht}}$ 
    scale is smaller by about 1.4 dex compared 
    to that of the  $\mstar-\mhalo$ relation . This suggests that baryonic processes like heating
    and feedback suppress the HI content in large mass halos on a shorter timescale 
    as compared to star-formation.
    
\end{itemize}

In this work we have obtained scaling relations of HI selected galaxies. The scaling relations 
among HI properties, i.e. $\mhi-\w50-\Vrothi$, have been obtained from 
the same sample with no model assumptions. Although figure~\ref{fig_scaling_compare} shows 
that our results are consistent with the data (and provides for a consistency check of the approach
taken in this work) we expect scatter to flatten these relations
and the relations between HI properties and halo mass at the high mass end
\citep{2010ApJ...717..379B,2020ApJ...894..124R}. In summary we have effectively constrained
a multivariate HI halo model, but unlike traditional approaches in constructing a halo model
we have not used HI clustering to constrain it, but have rather constrained  
the HI - halo mass scaling relations from the results of \citet{2020ApJ...894...92G}. In a forthcoming
paper we will present our clustering predictions of HI selected galaxies based on the current work.

\section*{ACKNOWLEDGMENTS}
We would like to thank the referee for comments and suggestions which have improved the presentation 
of the paper. We would like to thank  Raghunathan Srianand, Aseem Paranjape and Jasjeet Singh Bagla 
for useful discussions. SD would like to thank Somnath Bharadwaj for useful comments.
NK acknowledges the support of the Ramanujan 
Fellowship\footnote{Awarded by the 
Department of Science and Technology, Government of India} 
and  the IUCAA\footnote{Inter University Centre for Astronomy and 
Astrophysics, Pune, India} associateship programme. 
All the analyses were done on the {\tiny{\bf XANADU}} and {\tiny{\bf CHANDRA}} servers 
funded by the Ramanujan Fellowship. SD would like to thank Sudhakar Panda for financial 
support from the J.C. Bose Fellowship$^1$. 

We thank the entire ALFALFA collaboration 
in observing, flagging, and extracting the properties of galaxies that this paper 
makes use of. This work also uses data from SDSS DR7.
Funding for the SDSS and SDSS-II has been provided by the Alfred P. Sloan Foundation, 
the Participating Institutions, the National Science Foundation, 
the U.S. Department of Energy, the National Aeronautics and Space Administration, 
the Japanese Monbukagakusho, the Max Planck Society, and the Higher 
Education Funding Council for England. The SDSS Website is 
http://www.sdss.org/. The SDSS is managed by the Astrophysical Research 
Consortium for the Participating Institutions.

The CosmoSim database used in this paper is 
a service by the Leibniz-Institute for Astrophysics Potsdam (AIP).
The MultiDark database was developed in cooperation with 
the Spanish MultiDark Consolider Project CSD2009-00064.
The authors gratefully acknowledge the Gauss Centre for Supercomputing e.V. 
(www.gauss-centre.eu) and 
the Partnership for Advanced Supercomputing in Europe (PRACE, www.prace-ri.eu) 
for funding the MultiDark simulation project by providing computing time on 
the GCS Supercomputer SuperMUC at Leibniz Supercomputing Centre (LRZ, www.lrz.de).
The Bolshoi simulations have been performed within 
the Bolshoi project of the University of California 
High-Performance AstroComputing Center (UC-HiPACC) 
and were run at the NASA Ames Research Center.

\section*{Data Availability}
The data used in this work is publicly available.
SDSS DR7 \citep{2009ApJS..182..543A}
data can be accessed from \emph{sciserver.org}
and the $\alpha.40$ \citep{2011AJ....142..170H} 
data from ALFALFA can be accessed from 
\emph{egg.astro.cornell.edu}. The MultiDark simulations 
and galaxy catalogs are available in 
the CosmoSim database at \emph{https://www.cosmosim.org}.

\appendix
\section{The Effect of $\lowercase{w_{\text{nr}}}$ on the HIWF and HIVF}
\label{appendix}
\begin{figure*}
  \begin{tabular}{cc}
    \includegraphics[width=3.4in]{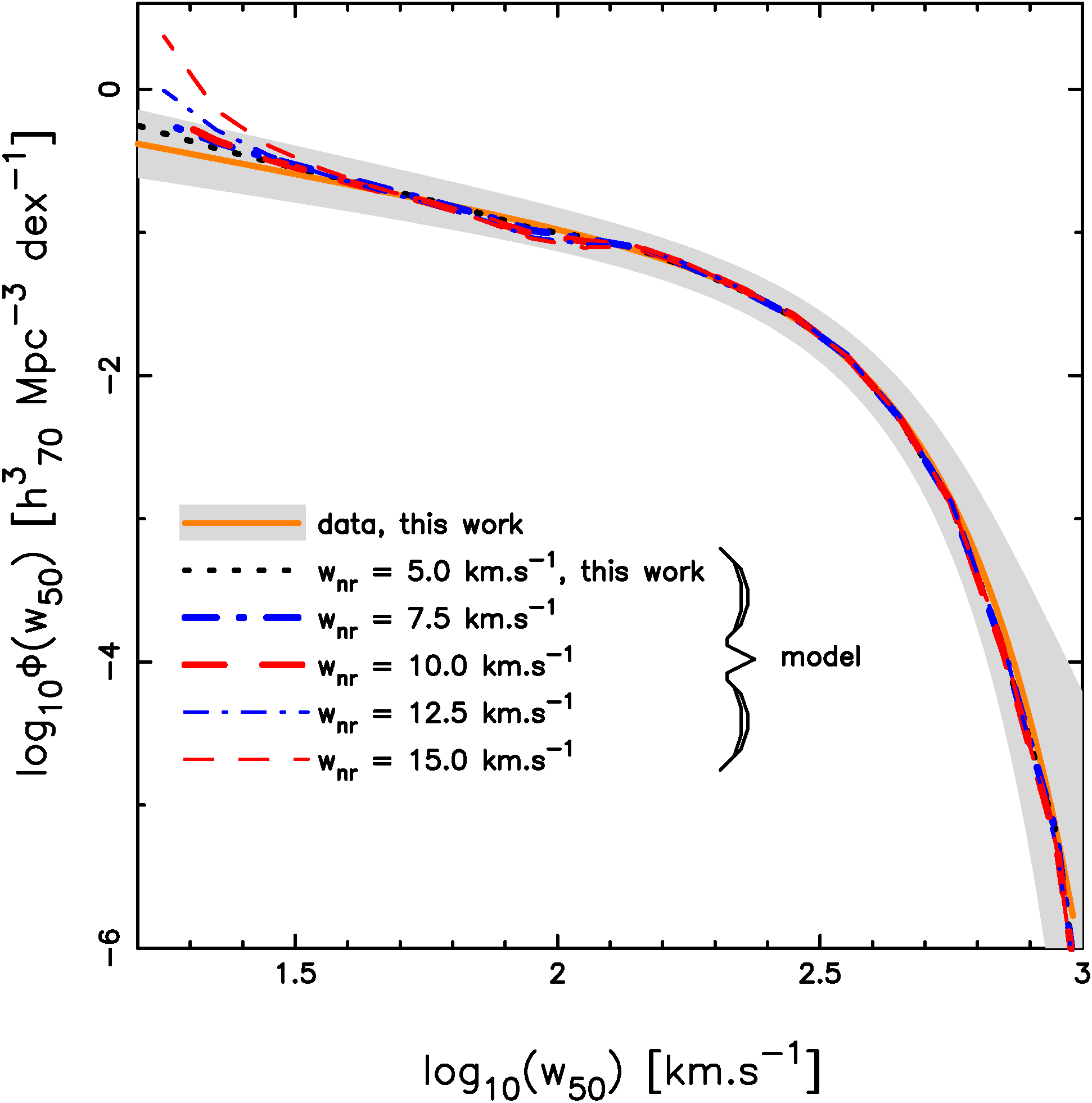}    
    \includegraphics[width=3.4in]{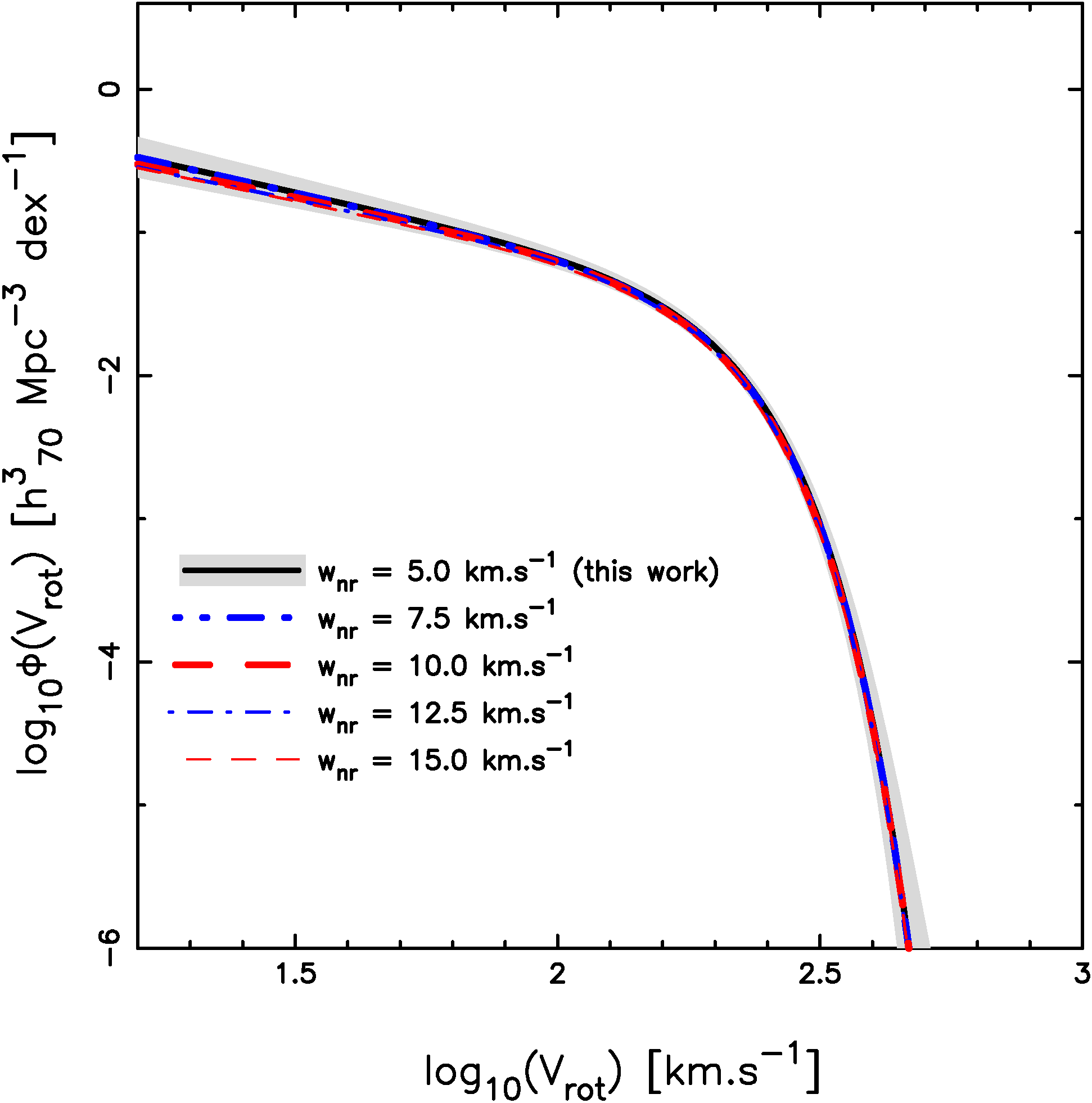}\\
  \end{tabular}
  \caption{This figure 
    shows the effect of varying  $w_{nr}$ on  the HIWF (HIVF) in the left (right)
    panel. In this work we considered $w_{nr} = 5.0 \; \text{km.s}^{-1}$ (see section~\ref{sec_hiwf}). 
    The solid line with the shaded region in the left panel represents the best fit 
    modified Schechter function for the HIWF and its $1\sigma$ uncertainty based on the data.
    The thick dotted line in the left panel represents the best fit model HIWF 
    for $w_{nr} = 5.0 \; \text{km.s}^{-1}$ which was determined by a corresponding 
    best fit  HIVF in the right panel (solid line with shaded region). 
    The thick dot-dashed, thick dashed, thin dot-dashed, thin dashed lines represent the best fit HIWF
    for $w_{nr} = 7.5 \; \text{km.s}^{-1}, 10.0 \; \text{km.s}^{-1}, 
    12.5 \; \text{km.s}^{-1}, 15.0 \; \text{km.s}^{-1}$ respectively, determined from 
    their corresponding best fit HIVF shown on the right panel.}
  \label{fig_hifn}
\end{figure*}

\begin{table}
\begin{center}
\resizebox{0.45\textwidth}{!}{
\begin{tabular}{|l|c|c|c|c|}
\hline
$w_{nr}$ & $\phi_*$  & $V_*$ & $\alpha$ & $\beta$ \\
\hline
5.0 & 0.0187 $\pm$ 0.0023 & 2.30 $\pm$ 0.01 & -0.81 $\pm$ 0.13 & 2.70 $\pm$ 0.17 \\
\hline
7.5 & 0.0182 $\pm$ 0.0023 & 2.30 $\pm$ 0.01 & -0.82 $\pm$ 0.13 & 2.70 $\pm$ 0.17 \\
\hline
10.0 & 0.0193 $\pm$ 0.0022 & 2.29 $\pm$ 0.01 & -0.76 $\pm$ 0.13 & 2.64 $\pm$ 0.17 \\ 
\hline
12.5 & 0.0189 $\pm$ 0.0022 & 2.29 $\pm$ 0.01 & -0.75 $\pm$ 0.13 & 2.65 $\pm$ 0.17 \\ 
\hline
15.0 & 0.0182 $\pm$ 0.0022 & 2.29 $\pm$ 0.01 & -0.76 $\pm$ 0.13 & 2.65 $\pm$ 0.17 \\ 
\hline
\end{tabular}
}
\caption{Best fit values with $1\sigma$ uncertainties of 
modified Schechter function for HIVFs for different values of $w_{nr}$.}
\label{tab_bf_app}
\end{center}
\end{table}
We look at the effect of varying $w_{\text{nr}}$ on the derived HIVF in this section. 
We proceed with the approach described in section~\ref{sec_hiwf}. 
We use equation~\ref{eq_w50_vrot} 
\citep{2001A&A...370..765V,2011ApJ...739...38P} to convert $\Vrot$ into $\w50$. 
We add $w_{\text{nr}}$ to $\Vrot$ linearly for galaxies 
with $\Vrot > 50 \text{km.s}^{-1}$ and in quadrature for galaxies with smaller velocities 
\citep{2011ApJ...739...38P}. The best fit model parameters of the HIVF are determined 
by a minimization procedure so that the corresponding model HIWF and the observed HIWF agree with each 
other. Our results are shown in figure~\ref{fig_hifn}. The right panel shows the best fit HIVF for 
values of $w_{\text{nr}}$ in the range $5 - 15 \text{km.s}^{-1}$ \citep{2013ApJ...773...88S}. The 
left panel shows the corresponding model HIWF and the HIWF from observations
(thick solid line with shaded region). We find negligible variation in the model HIVFs as a function 
of $w_{\text{nr}}$. This is also true for the model HIWFs for $\w50 \gsim 30 \text{km.s}^{-1} $.
For $\w50 < 30 \text{km.s}^{-1} $ there is an upturn in the model HIWFs for 
$w_{\text{nr}} \geq 12.5 \text{km.s}^{-1}$ which is not consistent with data.   

The parameters of the best fit HIVFs for different values of   $w_{\text{nr}}$ are shown 
in table~\ref{tab_bf_app}. Although the HIVFs are consistent with each other, the effect of varying 
$w_{\text{nr}}$ shows up most significantly at lower velocities, i.e. in the slope $\alpha$.

\label{lastpage}


\begin{thebibliography}{}


\bibitem[\protect\citeauthoryear{Abazajian, et al.}{2009}]{2009ApJS..182..543A}
  Abazajian, K. N., et al. 
  2009,ApJS,182, 543

\bibitem[\protect\citeauthoryear{Planck Collaboration, Ade et al.}{2016}]{2016A&A...594A..13P} 
  Planck Collaboration, Ade P.~A.~R., Aghanim N., Arnaud M., Ashdown M., Aumont J., et al., 
  2016, A\&A, 594, A13. 

\bibitem[\protect\citeauthoryear{Ahn et al.}{2014}]{2014ApJS..211...17A} 
  Ahn C.~P., Alexandroff R., Allende Prieto C., Anders F., Anderson S.~F., et al., 
  2014, ApJS, 211, 17. 

\bibitem[\protect\citeauthoryear{Bagla, Khandai, \& Datta}{2010}]{2010MNRAS.407..567B} 
  Bagla J.~S., Khandai N., Datta K.~K., 
  2010, MNRAS, 407, 567. 

\bibitem[\protect\citeauthoryear{Baldry, et al.}{2004}]{2004ApJ...600..681B}
  Baldry, I. K., et al. 
  2004, ApJ,600,681

\bibitem[\protect\citeauthoryear{Baldry, et al.}{2012}]{2012MNRAS.421..621B}
  Baldry et al., 
  2012, MNRAS, 421, 621

\bibitem[\protect\citeauthoryear{Behroozi, Conroy, \& Wechsler}{2010}]{2010ApJ...717..379B} 
  Behroozi P.~S., Conroy C., Wechsler R.~H., 
  2010, ApJ, 717, 379. 

\bibitem[\protect\citeauthoryear{Behroozi, Wechsler, \& Conroy}{2013}]{2013ApJ...770...57B} 
  Behroozi P.~S., Wechsler R.~H., Conroy C., 
  2013, ApJ, 770, 57. 

\bibitem[\protect\citeauthoryear{Behroozi, Wechsler, \& Wu}{2013}]{2013ApJ...762..109B} 
  Behroozi P.~S., Wechsler R.~H., Wu H.-Y., 
  2013, ApJ, 762, 109. 

\bibitem[\protect\citeauthoryear{Behroozi et al.}{2013}]{2013ApJ...763...18B} 
  Behroozi P.~S., Wechsler R.~H., Wu H.-Y., Busha M.~T., Klypin A.~A., Primack J.~R., 
  2013, ApJ, 763, 18. 

\bibitem[\protect\citeauthoryear{Behroozi et al.}{2019}]{2019MNRAS.488.3143B} 
  Behroozi P., Wechsler R.~H., Hearin A.~P., Conroy C., 
  2019, MNRAS, 488, 3143. 

\bibitem[\protect\citeauthoryear{Blanton \& Roweis}{2007}]{2007AJ....133..734B}
  Blanton M. R., Roweis S., 
  2007, AJ, 133, 734

\bibitem[\protect\citeauthoryear{Casey, Narayanan \& Cooray}{2014}]{2014PhR...541...45C} 
  Casey C.~M., Narayanan D., Cooray A., 
  2014, PhR, 541, 45

\bibitem[\protect\citeauthoryear{Carilli \& Walter}{2013}]{2013ARA&A..51..105C} 
  Carilli C.~L., Walter F., 
  2013, ARA\&A, 51, 105. 

\bibitem[\protect\citeauthoryear{Catinella, et al.}{2013}]{2013MNRAS.436...34C} 
  Catinella B., et al., 
  2013, MNRAS, 436, 34

\bibitem[\protect\citeauthoryear{Chae}{2010}]{2010MNRAS.402.2031C} 
  Chae K.-H., 
  2010, MNRAS, 402, 2031. 
  
\bibitem[\protect\citeauthoryear{Chaves-Montero et al.}{2016}]{2016MNRAS.460.3100C} 
  Chaves-Montero J., Angulo R.~E., Schaye J., Schaller M., Crain R.~A., Furlong M., Theuns T., 
  2016, MNRAS, 460, 3100. 


\bibitem[\protect\citeauthoryear{Conroy \& Wechsler}{2009}]{2009ApJ...696..620C} 
  Conroy C., Wechsler R.~H., 
  2009, ApJ, 696, 620. 

\bibitem[\protect\citeauthoryear{Conroy, Gunn, \& White}{2009}]{2009ApJ...699..486C} 
  Conroy C., Gunn J.~E., White M., 
  2009, ApJ, 699, 486. 

\bibitem[\protect\citeauthoryear{Courtois et al.}{2009}]{2009AJ....138.1938C} 
  Courtois H.~M., Tully R.~B., Fisher J.~R., Bonhomme N., Zavodny M., Barnes A., 
  2009, AJ, 138, 1938. 

\bibitem[\protect\citeauthoryear{Crain et al.}{2015}]{2015MNRAS.450.1937C} 
  Crain R.~A., Schaye J., Bower R.~G., Furlong M., Schaller M., et al., 
  2015, MNRAS, 450, 1937. 

\bibitem[\protect\citeauthoryear{Crain et al.}{2017}]{2017MNRAS.464.4204C} 
  Crain R.~A., Bah{\'e} Y.~M., Lagos C. del P., Rahmati A., Schaye J., et al., 
  2017, MNRAS, 464, 4204. 

\bibitem[\protect\citeauthoryear{Dav{\'e} et al.}{2019}]{2019MNRAS.486.2827D} 
  Dav{\'e} R., Angl{\'e}s-Alc{\'a}zar D., Narayanan D., Li Q., Rafieferantsoa M.~H., Appleby S., 
  2019, MNRAS, 486, 2827. 

\bibitem[\protect\citeauthoryear{Dav{\'e} et al.}{2020}]{2020MNRAS.497..146D} 
  Dav{\'e} R., Crain R.~A., Stevens A.~R.~H., Narayanan D., Saintonge A., et al., 
  2020, MNRAS, 497, 146. 

\bibitem[\protect\citeauthoryear{Davis \& Huchra}{1982}]{1982ApJ...254..437D} 
  Davis M., Huchra J., 
  1982, ApJ, 254, 437 

\bibitem[\protect\citeauthoryear{Diemer et al.}{2019}]{2019MNRAS.487.1529D} 
  Diemer B., Stevens A.~R.~H., Lagos C. del P., Calette A.~R., Tacchella S., et al., 
  2019, MNRAS, 487, 1529. 

\bibitem[\protect\citeauthoryear{Di Matteo et al.}{2012}]{2012ApJ...745L..29D} 
  Di Matteo T., Khandai N., DeGraf C., Feng Y., Croft R.~A.~C., et al., 
  2012, ApJL, 745, L29. 

\bibitem[\protect\citeauthoryear{Drory, et al.}{2009}]{2009ApJ...707.1595D} 
  Drory N., et al., 
  2009, ApJ, 707, 1595

\bibitem[\protect\citeauthoryear{Dutta, Khandai, \& Dey}{2020}]{2020MNRAS.494.2664D} 
  Dutta S., Khandai N., Dey B., (D20) 
  2020, MNRAS, 494, 2664

\bibitem[\protect\citeauthoryear{Dutta \& Khandai}{2021}]{2021MNRAS.500L..37D} 
  Dutta S., Khandai N., (D21)
  2021, MNRAS, 500, L37 

\bibitem[\protect\citeauthoryear{Durbala et al.}{2020}]{2020AJ....160..271D} 
  Durbala A., Finn R.~A., Crone Odekon M., Haynes M.~P., Koopmann R.~A., O'Donoghue A.~A., 
  2020, AJ, 160, 271. 

\bibitem[\protect\citeauthoryear{Efstathiou, Ellis, \& Peterson}{1988}]{1988MNRAS.232..431E} 
  Efstathiou G., Ellis R.~S., Peterson B.~A., 
  1988, MNRAS, 232, 431 

\bibitem[\protect\citeauthoryear{Feng et al.}{2016}]{2016MNRAS.455.2778F} 
  Feng Y., Di-Matteo T., Croft R.~A., Bird S., Battaglia N., et al.,  
  2016, MNRAS, 455, 2778. 
  
\bibitem[\protect\citeauthoryear{Gordon}{1971}]{1971ApJ...169..235G} 
  Gordon K.~J., 1971, ApJ, 169, 235. 

\bibitem[\protect\citeauthoryear{Guo et al.}{2017}]{2017ApJ...846...61G} 
  Guo H., Li C., Zheng Z., Mo H.~J., Jing Y.~P., et al., 
  2017, ApJ, 846, 61. 

\bibitem[\protect\citeauthoryear{Guo et al.}{2020}]{2020ApJ...894...92G} 
  Guo H., Jones M.G., Haynes M.P., Fu J., 
  2020, ApJ, 894, 92

\bibitem[\protect\citeauthoryear{Haynes, et al.}{2011}]{2011AJ....142..170H}
  Haynes M. P., et al., 
  2011, AJ, 142, 170

\bibitem[\protect\citeauthoryear{Haynes, et al.}{2018}]{2018ApJ...861...49H}
  Haynes M. P., et al., 
  2018 ApJ 861, 49

\bibitem[\protect\citeauthoryear{Huang, et al.}{2012}]{2012ApJ...756..113H} 
  Huang S., Haynes M.~P., Giovanelli R., Brinchmann J., 
  2012, ApJ, 756, 113

\bibitem[\protect\citeauthoryear{Kennicutt}{1989}]{1989ApJ...344..685K} 
  Kennicutt R.~C., 
  1989, ApJ, 344, 685

\bibitem[\protect\citeauthoryear{Kennicutt}{1998}]{1998ApJ...498..541K} 
  Kennicutt R.~C., 
  1998, ApJ, 498, 541


\bibitem[\protect\citeauthoryear{Kim, et al.}{2017}]{2017MNRAS.465..111K} 
  Kim H.-S. et al., 
  2017, MNRAS, 465, 111

\bibitem[\protect\citeauthoryear{Khandai, et al.}{2011}]{2011MNRAS.415.2580K} 
  Khandai N., et al., 
  2011, MNRAS, 415, 2580 

\bibitem[\protect\citeauthoryear{Khandai et al.}{2015}]{2015MNRAS.450.1349K} 
  Khandai N., Di Matteo T., Croft R., Wilkins S., Feng Y., et al., 
  2015, MNRAS, 450, 1349. 

\bibitem[\protect\citeauthoryear{Klypin et al.}{2016}]{2016MNRAS.457.4340K} 
  Klypin A., Yepes G., Gottl{\"o}ber S., Prada F., He{\ss} S., 
  2016, MNRAS, 457, 4340. 

\bibitem[\protect\citeauthoryear{Knebe et al.}{2018}]{2018MNRAS.474.5206K} 
  Knebe A., Stoppacher D., Prada F., Behrens C., Benson A., Cora S.~A., Croton D.~J., et al., 
  2018, MNRAS, 474, 5206. 

\bibitem[\protect\citeauthoryear{Lah, et al.}{2009}]{2009MNRAS.399.1447L} 
  Lah P., et al., 
  2009, MNRAS, 399, 1447

\bibitem[\protect\citeauthoryear{Lelli, McGaugh, \& Schombert}{2016}]{2016AJ....152..157L} 
  Lelli F., McGaugh S.~S., Schombert J.~M., 
  2016, AJ, 152, 157. 

\bibitem[\protect\citeauthoryear{Lelli et al.}{2017}]{2017ApJ...836..152L} 
  Lelli F., McGaugh S.~S., Schombert J.~M., Pawlowski M.~S., 
  2017, ApJ, 836, 152. 

\bibitem[\protect\citeauthoryear{Lemonias et al.}{2013}]{2013ApJ...776...74L} 
  Lemonias J.~J., Schiminovich D., Catinella B., Heckman T.~M., Moran S.~M., 
  2013, ApJ, 776, 74

\bibitem[\protect\citeauthoryear{Li et al.}{2019a}]{2019ApJ...886L..11L} 
  Li P., Lelli F., McGaugh S., Pawlowski M.~S., Zwaan M.~A., Schombert J., 
  2019, ApJL, 886, L11. 

\bibitem[\protect\citeauthoryear{Li et al.}{2019b}]{2019MNRAS.482.5106L} 
  Li P., Lelli F., McGaugh S.~S., Starkman N., Schombert J.~M., 
  2019, MNRAS, 482, 5106. 
  

\bibitem[\protect\citeauthoryear{Li et al.}{2020}]{2020ApJS..247...31L} 
  Li P., Lelli F., McGaugh S., Schombert J., 
  2020, ApJS, 247, 31. 


\bibitem[\protect\citeauthoryear{Loveday}{2000}]{2000MNRAS.312..557L}
  Loveday, J., 
  2000, MNRAS,312,557

\bibitem[\protect\citeauthoryear{Dutton \& Macci{\`o}}{2014}]{2014MNRAS.441.3359D} 
  Dutton A.~A., Macci{\`o} A.~V., 
  2014, MNRAS, 441, 3359. 

\bibitem[\protect\citeauthoryear{Moster, Naab, \& White}{2013}]{2013MNRAS.428.3121M} 
  Moster B.~P., Naab T., White S.~D.~M., 
  2013, MNRAS, 428, 3121. 


\bibitem[\protect\citeauthoryear{Navarro, Frenk, \& White}{1996}]{1996ApJ...462..563N} 
  Navarro J.~F., Frenk C.~S., White S.~D.~M., 
  1996, ApJ, 462, 563. 


\bibitem[\protect\citeauthoryear{Madau \& Dickinson}{2014}]{2014ARA&A..52..415M} 
  Madau P., Dickinson M., 
  2014, ARA\&A, 52, 415. 


\bibitem[\protect\citeauthoryear{Maddox, et al.}{2015}]{2015MNRAS.447.1610M} 
  Maddox N., Hess K.~M., Obreschkow D., Jarvis M.~J., Blyth S.-L., 
  2015, MNRAS, 447, 1610

\bibitem[\protect\citeauthoryear{Martin, et al.}{2010}]{2010ApJ...723.1359M}
  Martin, A., et al., 
  2010, ApJ, 723,1359


\bibitem[\protect\citeauthoryear{Meyer et al.}{2004}]{2004MNRAS.350.1195M} 
  Meyer M.~J., Zwaan M.~A., Webster R.~L., Staveley-Smith L., Ryan-Weber E., et al., 
  2004, MNRAS, 350, 1195. 

\bibitem[\protect\citeauthoryear{Moorman, et al.}{2014}]{2014MNRAS.444.3559M} 
  Moorman C.~M., et al., 
  2014, MNRAS, 444, 3559 

\bibitem[\protect\citeauthoryear{Moustakas et al.}{2013}]{2013ApJ...767...50M} 
  Moustakas J., Coil A.~L., Aird J., Blanton M.~R., Cool R.~J., Eisenstein D.~J., Mendez A.~J., et al.,
  2013, ApJ, 767, 50. 

\bibitem[\protect\citeauthoryear{Naab \& Ostriker}{2017}]{2017ARA&A..55...59N} 
  Naab T., Ostriker J.~P., 
  2017, ARA\&A, 55, 59. 

\bibitem[\protect\citeauthoryear{Obuljen et al.}{2019}]{2019MNRAS.486.5124O} 
  Obuljen A., Alonso D., Villaescusa-Navarro F., Yoon I., Jones M., 
  2019, MNRAS, 486, 5124. 

\bibitem[\protect\citeauthoryear{Padmanabhan \& Kulkarni}{2017}]{2017MNRAS.470..340P} 
  Padmanabhan H., Kulkarni G., 
  2017, MNRAS, 470, 340

\bibitem[\protect\citeauthoryear{Padmanabhan, Refregier, \& Amara}{2017}]{2017MNRAS.469.2323P} 
  Padmanabhan H., Refregier A., Amara A., 
  2017, MNRAS, 469, 2323. 

\bibitem[\protect\citeauthoryear{Papastergis, et al.}{2011}]{2011ApJ...739...38P} 
  Papastergis, E., Martin, A.M., Giovanelli, R. \& Haynes, M.P. 
  2010, ApJ 739, 38

\bibitem[\protect\citeauthoryear{Paranjape, Choudhury, \& Sheth}{2021}]{2021MNRAS.503.4147P} 
  Paranjape A., Choudhury T.~R., Sheth R.~K., 
  2021, MNRAS, 503, 4147. 

\bibitem[\protect\citeauthoryear{Paranjape et al.}{2021}]{2021arXiv210504570P} 
  Paranjape A., Srianand R., Choudhury T.~R., Sheth R.~K., 
  2021, arXiv, arXiv:2105.04570

\bibitem[\protect\citeauthoryear{Paul, Choudhury, \& Paranjape}{2018}]{2018MNRAS.479.1627P} 
  Paul N., Choudhury T.~R., Paranjape A., 
  2018, MNRAS, 479, 1627. 

\bibitem[\protect\citeauthoryear{Pillepich et al.}{2018}]{2018MNRAS.475..648P} 
  Pillepich A., Nelson D., Hernquist L., Springel V., Pakmor R., et al., 
  2018, MNRAS, 475, 648. 

\bibitem[\protect\citeauthoryear{Rahmati et al.}{2013}]{2013MNRAS.431.2261R} 
  Rahmati A., Schaye J., Pawlik A.~H., Rai{\v{c}}evi{\'c} M., 
  2013, MNRAS, 431, 2261. 
  

\bibitem[\protect\citeauthoryear{Ren, Trenti, \& Di Matteo}{2020}]{2020ApJ...894..124R} 
  Ren K., Trenti M., Di Matteo T., 
  2020, ApJ, 894, 124. 

\bibitem[\protect\citeauthoryear{Rhee, et al.}{2018}]{2018MNRAS.473.1879R} 
  Rhee J., et al., 
  2018, MNRAS, 473, 1879
  
\bibitem[\protect\citeauthoryear{Romeo}{2020}]{2020MNRAS.491.4843R}
  Romeo A.~B., 2020, MNRAS, 491, 4843. 
  
\bibitem[\protect\citeauthoryear{Romeo, Agertz, \& Renaud}{2020}]{2020MNRAS.499.5656R} 
  Romeo A.~B., Agertz O., Renaud F., 2020, MNRAS, 499, 5656. 

\bibitem[\protect\citeauthoryear{Schaye et al.}{2015}]{2015MNRAS.446..521S} 
  Schaye J., Crain R.~A., Bower R.~G., Furlong M., Schaller M., et al., 
  2015, MNRAS, 446, 521. 

\bibitem[\protect\citeauthoryear{Schmidt}{1959}]{1959ApJ...129..243S} 
  Schmidt M., 
  1959, ApJ, 129, 243

\bibitem[\protect\citeauthoryear{Schmidt}{1963}]{1963ApJ...137..758S} 
  Schmidt M., 
  1963, ApJ, 137, 758
  
\bibitem[\protect\citeauthoryear{Schulman, Bregman, \& Roberts}{1994}]{1994ApJ...423..180S}
  Schulman E., Bregman J.~N., Roberts M.~S., 1994, ApJ, 423, 180. 

\bibitem[\protect\citeauthoryear{Serra et al.}{2012}]{2012MNRAS.422.1835S} 
  Serra P., Oosterloo T., Morganti R., Alatalo K., Blitz L., Bois M., Bournaud F., et al., 
  2012, MNRAS, 422, 1835. 

\bibitem[\protect\citeauthoryear{Somerville \& Dav{\'e}}{2015}]{2015ARA&A..53...51S} 
  Somerville R.~S., Dav{\'e} R., 
  2015, ARA\&A, 53, 51. 

\bibitem[\protect\citeauthoryear{Spinelli et al.}{2020}]{2020MNRAS.493.5434S} 
  Spinelli M., Zoldan A., De Lucia G., Xie L., Viel M., 
  2020, MNRAS, 493, 5434. 

\bibitem[\protect\citeauthoryear{Springel \& Hernquist}{2003}]{2003MNRAS.339..289S}
  Springel V., Hernquist L., 
  2003, MNRAS, 339, 289

\bibitem[\protect\citeauthoryear{Springel}{2005}]{2005MNRAS.364.1105S} 
  Springel V., 
  2005, MNRAS, 364, 1105. 

\bibitem[\protect\citeauthoryear{Stevens et al.}{2021}]{2021MNRAS.502.3158S} 
  Stevens A.~R.~H., Lagos C. del P., Cortese L., Catinella B., Diemer B., et al., 
  2021, MNRAS, 502, 3158. 
  
\bibitem[\protect\citeauthoryear{Stilp et al.}{2013}]{2013ApJ...773...88S} 
  Stilp A.~M., Dalcanton J.~J., Skillman E., Warren S.~R., Ott J., Koribalski B., 
  2013, ApJ, 773, 88. 
  
\bibitem[\protect\citeauthoryear{Stiskalek et al.}{2021}]{2021MNRAS.506.3205S} 
  Stiskalek R., Desmond H., Holvey T., Jones M.~G., 2021, MNRAS, 506, 3205. 

\bibitem[\protect\citeauthoryear{Tully et al.}{2009}]{2009AJ....138..323T} 
  Tully R.~B., Rizzi L., Shaya E.~J., Courtois H.~M., Makarov D.~I., Jacobs B.~A., 
  2009, AJ, 138, 323. 
  
\bibitem[\protect\citeauthoryear{Vale \& Ostriker}{2004}]{2004MNRAS.353..189V} 
  Vale A., Ostriker J.~P., 2004, MNRAS, 353, 189. 

\bibitem[\protect\citeauthoryear{Verheijen \& Sancisi}{2001}]{2001A&A...370..765V} 
  Verheijen M.~A.~W., Sancisi R., 
  2001, A\&A, 370, 765. 

\bibitem[\protect\citeauthoryear{Verheijen}{2001}]{2001ApJ...563..694V} 
  Verheijen M.~A.~W., 
  2001, ApJ, 563, 694. 

\bibitem[\protect\citeauthoryear{Villaescusa-Navarro et al.}{2018}]{2018ApJ...866..135V} 
  Villaescusa-Navarro F., Genel S., Castorina E., Obuljen A., Spergel D.~N., et al., 
  2018, ApJ, 866, 135. 

\bibitem[\protect\citeauthoryear{Vogelsberger et al.}{2014}]{2014Natur.509..177V} 
  Vogelsberger M., Genel S., Springel V., Torrey P., Sijacki D., et al., 
  2014, Nature, 509, 177. 

\bibitem[\protect\citeauthoryear{Wang et al.}{2016}]{2016MNRAS.460.2143W} 
  Wang J., Koribalski B.~S., Serra P., van der Hulst T., Roychowdhury S., Kamphuis P., 
  Chengalur J.~N., 
  2016, MNRAS, 460, 2143. 

\bibitem[\protect\citeauthoryear{Zwaan, Briggs, \& Sprayberry}{2001}]{2001MNRAS.327.1249Z} 
  Zwaan M.~A., Briggs F.~H., Sprayberry D., 
  2001, MNRAS, 327, 1249

\bibitem[\protect\citeauthoryear{Zwaan, et al.}{2003}]{2003AJ....125.2842Z} 
  Zwaan M.~A., et al., 
  2003, AJ, 125, 2842 

\bibitem[\protect\citeauthoryear{Zwaan, et al.}{2010}]{2010MNRAS.403.1969Z} 
  Zwaan, M. A.; Meyer, M. J.; Staveley-Smith, L., 
  2010, MNRAS, Volume 403, Issue 4, pp. 1969-1977

\end{thebibliography}
\end{document}